\documentclass[12pt]{article}
\pdfoutput=1

\usepackage{draft,hyperref,tikz,cancel,subfig,float}
\usepackage[nosort]{cite}

\usetikzlibrary{snakes}
\usetikzlibrary{shapes.misc}
\usepackage{todonotes}
\usepackage{framed}
\usepackage{slashed}
\begin{document}

\begin{titlepage}

\vspace*{-2.5cm} 
\noindent\hfill USTC-ICTS/PCFT-26-24\\\\

\begin{center}
\title{Classification of 2D Fermionic Systems with a $\mathbb{Z}_2$ Flavor Symmetry}
\author{Chi-Ming Chang$^{a,b}$,  Jin Chen$^{c,d}$,  and Fengjun Xu$^{b}$}
\address{\small${}^a$Yau Mathematical Sciences Center (YMSC), Tsinghua University, Beijing, 100084, China}
\address{\small${}^b$Beijing Institute of Mathematical Sciences and Applications (BIMSA), Beijing, 101408, China}

\address{\small${}^c$Department of Physics, Xiamen University, Xiamen, 361005, China}

\address{\small${}^d$Peng Huanwu Center for Fundamental Theory, Hefei, Anhui 230026, China}
%
\email{cmchang@tsinghua.edu.cn, zenofox@gmail.com, xufengjun321@gmail.com}
\end{center}
\vfill
\begin{abstract}
We classify superfusion categories describing two-dimensional fermionic systems equipped with the universal fermion-parity symmetry, implemented by a topological defect line (TDL) 
$Z$, and an additional $\mathbb{Z}_2$ flavor symmetry generated by a $W$ TDL. Depending on whether $W$ is m-type or q-type, its fusion rules lead to three distinct classes, and solving the super-pentagon equations yields 16 consistent superfusion categories. These are labeled by invariants $(\nu_W,\nu_Z,\nu_{WZ})$, which determine the $\mathbb{Z}_8$ anomaly classes of the symmetries generated by $W$, $Z$, and $WZ$. We also provide explicit realizations using multiple Majorana fermions and comment on implications for fermionic CFTs and gapped phases.
\end{abstract}
\vfill
\end{titlepage}

\tableofcontents

\section{Introduction}
In recent years, the framework for describing symmetries in quantum field theory has undergone significant development. From a modern perspective, symmetries in a QFT can be characterized by the presence of topological defect operators \cite{Gaiotto:2014kfa}. In a two-dimensional quantum field theory, particularly conformal field theory, such topological defect operators are known as topological defect lines (TDLs). TDLs generalize the familiar group-like (invertible) symmetries to encompass non-invertible symmetries \cite{Frohlich:2004ef, Frohlich:2006ch, Frohlich:2009gb}, with both types organized uniformly within the mathematical framework of fusion categories; see \cite{Bhardwaj:2017xup, Chang:2018iay} for recent developments.

When a two-dimensional CFT carries fermionic degrees of freedom, then TDLs come in two types: m-type and q-type\cite{Gu:2010na, Gu:2013azn, Gaiotto:2015zta}. The distinction lies in the fact that q-type defects, in contrast to m-type defects, can support one-dimensional Majorana fermion modes along the defect. With such, it has important consequences for fusion, duality, and locality, and cannot be adequately captured within the framework of ordinary (bosonic) fusion categories. Instead, the appropriate mathematical structure for organizing the fusion and symmetry properties of defect lines in fermionic theories is provided by superfusion categories \cite{Brundan_2017, usher2018fermionic}, which incorporate fermion parity grading and allow for a systematic treatment of both m-type and q-type objects \cite{Aasen:2017ubm, Chang:2022hud}. From the symmetry perspective, superfusion categories play a role analogous to that of fusion categories in bosonic theories: they provide a unified framework that accommodates both invertible and non-invertible symmetries, while faithfully encoding the intrinsic fermionic nature of the theory. In particular, they make it possible to describe generalized symmetries in fermionic CFTs in a way that is compatible with spin structure, fermion parity, and the presence of Majorana modes on defects. There are many recent developments along this direction; see e.g. \cite{Komargodski:2020mxz,Bhardwaj:2024ydc,Inamura:2022lun, Kawabata:2024hzx, Wan:2016php, Lou:2020gfq, Ambrosino:2024ggh, Fukusumi:2024ejk, Fukusumi:2025ljx, Chen:2025qub} and references therein.

A key structural feature of any fermionic (conformal) field theory is the existence of a universal $\mathbb{Z}_2$ m-type topological defect line, denoted by $Z$, corresponding to the fermion parity symmetry $(-1)^F$.
This defect is present independently of any additional global symmetries and plays a distinguished role in the fusion algebra. Consequently, a complete characterization of the symmetry structure of a fermionic CFT must incorporate this universal $\mathbb{Z}_2$ symmetry as an intrinsic part of the associated superfusion category. Moreover, a systematic study of TDL-twisted partition functions associated with this $\mathbb{Z}_2$ symmetry has revealed intriguing connections to the spectral statistics of pure gravity in AdS$_3$ \cite{Boruch:2026hbr}.

In this note, we take a first step toward this goal by classifying superfusion category solutions describing fermionic systems endowed with an additional $\mathbb{Z}_2$ global (flavor) symmetry. In the convention of \cite{Chang:2022hud}, this additional global $\mathbb{Z}_2$ symmetry, characterized by the TDL $W$, falls into the following three distinct classes:
\begin{itemize}
    \item{ $ C_q^0:$ This corresponds to the so-called q-type, whose fusion ring with itself reads
\begin{equation}
    W^2=(1_b+1_f)I
\end{equation}}
The super-pentagon equations admit 4 solutions.

\item{$ C_m^{0,0}: $ This corresponds to the usual $m$-type $\mathbb{Z}_2$
\begin{equation}
    W^2=1_bI
\end{equation}}
The super-pentagon equations admit 2 solutions.  

\item{$ \hat{C}_m^{0,0}: $ This is also a $m$-type, but the difference with $ C_m^{0,0}$ is that the three junction $V_{W,W,I}$ is fermionic, i.e.
\begin{equation}
    W^2=1_fI
\end{equation}}
The super-pentagon equations admit 2 solutions. 
\end{itemize}
In total, there are 8 distinct superfusion categories, which are also denoted by $\mathbb{Z}_2^{\nu_W}$ for $\nu_w\in \mathbb{Z}_8$. The $C_q^0$, $C_m^{0,0}$, and $\hat{C}_m^{0,0}$ correspond to the $\mathbb{Z}_2^{\nu_W=1,3,5,7}$, $\mathbb{Z}_2^{\nu_W=0,4}$, and $\mathbb{Z}_2^{\nu_W=2,6}$, respectively. We briefly review this $\mathbb{Z}_8$ classification in Section~\ref{sec:z8cal}.

The universal $\mathbb{Z}_2$ TDL $Z$ (the fermion parity symmetry) should commute with the $\mathbb{Z}_2$ TDL $W$. In terms of their fusion ring, we have
\ie\label{eqn:fusion_ring}
W^2=(a 1_{\rm b}+b1_{\rm f})I,,\quad Z^2 = 1_{\rm b} I\,,\quad Z\,W=W\,Z\,,
\fe
where $a, b$ can take values in $\{0,1\}$ depending on which of the above three $Z_2$ class it takes. Hence, the classification of a $\mathbb{Z}_2$ global symmetry in a 2d fermionic system amounts to finding all consistent solutions to the super-pentagon equations for the corresponding superfusion categories with the fusion ring \eqref{eqn:fusion_ring}. In Section~\ref{sec:super-pentagon}, we solve these super-pentagon equations, and find that the solutions are classified by three parameters $\nu_W$, $\nu_Z$, and $\nu_{WZ}$. These parameters also specify the $\mathbb{Z}_8$ classification of the three $\mathbb{Z}_2$ symmetries generated by $W$, $Z$ and $WZ$. When $W$ is q-type, we find a new constraint on $\nu_W$ and $\nu_{WZ}$ from the anti-commutativity between the TDL $Z$ and the 1d Majorana fermion living on the TDL $W$.

In Sec.\ref{sec:8_Maj}, we construct explicit realizations of a subset of these solutions using $v$ copies of Majorana models. Finally, we discuss additional applications, including implications for gapped systems \ref{sec:app}.  The appendices contain supplementary material: Appendix \ref{app: F-symbols} collects the F-symbols associated with the three $\mathbb{Z}_2$ class solved in \cite{Chang:2022hud}. Appendix \ref{appen:gauging} presents the fermion action rules in $\mathbb{Z}_2^{\nu_W=1,3,5,7}\times \mathbb{Z}_2^{\nu_W=0,4}$. Appendix \ref{appen:2dMF} provides a short review on the Majorana fermions and establish a relation between the Kramers-Wannier duality line in the bosonic Ising and various TDLs in the fermionized one on the level of the partition functions. Appendix \ref{appix:fermionnumber} summarizes basic aspects of fermionic number operators in two-dimensional supersymmetric field theories relevant for \ref{sec:app}.

\section{A $\mathbb Z_8$ classification of the $\mathbb Z_2$ flavor symmetry}
\label{sec:z8cal}
Because $\mathbb Z_2$ symmetry constitutes a basic building block of our overall classification, it is useful to first review its classification in fermionic systems. In this section, we briefly recall the resulting $\mathbb Z_8$ classification of $\mathbb Z_2$ symmetry. 

A fermionic system is defined by specifying its spin structures, and a $\mathbb Z_2$ topological defect line is classified by the spin-cobordism group
\begin{align}
    \Omega_{2}^{\rm Spin}\left(B\mathbb Z_2,\,{\rm U}(1)\right)\simeq\mathbb Z_8\,.
\end{align}
This well-known result \cite{Gu:2013azn,Kapustin:2014dxa,Wang_2017, brumfiel2018pontrjagindual3dimensionalspin} can be also interpreted in terms of the super-pentagon solutions of $\mathbb Z_{2}$-symmetry \cite{Chang:2022hud}. In addition, realizations of these eight different $\mathbb Z_2$-lines are discussed in Sec.~\ref{sec:8_Maj}, (see also in \cite{Chang:2022hud}).

Let us denote the $\mathbb Z_2$-line by $W$, which is associated to the group element $\nu_W\in  \Omega_{2}^{\rm Spin}\left(B\mathbb Z_2\right)$. For $\nu_W=0,2,4,6$, the defect line is of m-type, whereas it is a q-type one for $\nu_W=1,3,5,7$. The parameter $\nu_W$ is related to the spin of the states in the defect Hilbert space ${\cal H}_W$ by
\ie
s_W={\rm spin}({\cal H}_W)=\frac{\nu_W}{16}\mod ~\frac12\,.
\fe
\paragraph{m-type:}
A m-type simple line $W$ has homomorphism
\begin{align}
    {\rm Hom}_{\mathbb Z_2^{{\rm even}\, \nu_W}}(W,\,W)\simeq \mathbb C^{1|0}\,,
\end{align}
where the $p$ and $q$ in the notation $\mathbb C^{p|q}$ denotes that the Hom space is $\mathbb Z_2$-graded and contains $p$/$q$-dimensional bosonic/fermionic space respectively. Although the m-type line has bosonic Hom space, the 3-ways junction could be either bosonic or fermionic \cite{Chang:2022hud}. More specifically, we have
\begin{align}
    {\rm Hom}_{\mathbb Z_2^{\nu_W}}(W\otimes W,\,I)\simeq\mathbb C^{1|0}\,,\quad {\rm for}\quad \nu_W=0,4\,,\notag\\
    {\rm Hom}_{\mathbb Z_2^{\nu_W}}(W\otimes W,\,I)\simeq\mathbb C^{0|1}\,,\quad {\rm for}\quad \nu_W=2,6\,.
\end{align}
Assigning the junction type to be either bosonic or fermionic, one can solve the super-pentagons and find two solutions, characterized by the Frobenius-Shur indicator $\kappa=\pm 1$, for a given junction type \cite{Chang:2022hud}. Therefore overall there are four solutions corresponding to the four even $\nu_W$ values, summarized in the table below:
\begin{table}[h!]
    \centering
    \begin{tabular}{c||c|c|c}
         &  Junction type & FS indicator & Spin selection rule $s=\frac{\nu}{16}\,({\rm mod\,\frac{1}{2}})$\\
         \hline
       $\nu_W=0$  &  $\mathbb C^{1|0}$ & $\kappa=+1$ & $s_W=0\,({\rm mod\,\frac{1}{2}})$\\
       \hline
       $\nu_W=2$  &  $\mathbb C^{0|1}$ & $\kappa=+1$ & $s_W=\frac{1}{8}\,({\rm mod\,\frac{1}{2}})$\\
       \hline
       $\nu_W=4$  &  $\mathbb C^{1|0}$ & $\kappa=-1$ & $s_W=\frac{1}{4}\,({\rm mod\,\frac{1}{2}})$\\
       \hline
       $\nu_W=6$  &  $\mathbb C^{0|1}$ & $\kappa=-1$ & $s_W=\frac{3}{8}\,({\rm mod\,\frac{1}{2}})$\\\hline
    \end{tabular}
    \caption{The classification of m-type $\mathbb{Z}_2$ TDLs}
    \label{tab:mtable}
\end{table}

\paragraph{q-type:}
A q-type simple line $W$ has homomorphism
\begin{align}
    {\rm Hom}_{\mathbb Z_2^{{\rm odd}\, \nu_W}}(W,\,W)\simeq \mathbb C^{1|1}\,,
\end{align}
It can be interpreted as that there is an 1D Majorana fermion living on the defect, because the Hilbert space of it in NS-sector after quantization is exactly isomorphic to $\mathbb C^{1|1}$. A direct consequence of this property is that the 3-ways junction is enforced to be 
\begin{align}
    {\rm Hom}_{\mathbb Z_2^{\nu_W}}(W\otimes W,\,I)\simeq\mathbb C^{1|1}\,,\quad {\rm for} \quad \nu_W =1,\,3,\,5,\,7\,.
    \label{eq:q_junc}
\end{align}
Assuming the junction type given in \eqref{eq:q_junc}, one can find four solutions to the super-pentagons \cite{Chang:2022hud}. The four solutions are classified by two gauge invariant parameters: the Frobenius-Shur indicator $\kappa=\pm 1$ and $\gamma=e^{\pi i/4}$ or $e^{3\pi i/4}$. The latter corresponds an additional phase by moving the 1D Majorana fermion across the 3-way junction in the following configuration:
\ie
\begin{gathered}
\begin{tikzpicture}[scale=1]
\draw [ultra thick, blue, densely dotted] (0,0)  -- (0,-1) ;
\draw [ultra thick, red, -<-=.6] (0,0) -- (.87,.5) ;
\draw [ultra thick, red, -<-=.6] (0,0) -- (-.87,.5) ;
\draw (0,0)\dotsolb {right}{};
\node at (-1, 0.7) {\tiny{$W$}};
\node at (1, 0.7) {\tiny{$W$}};
\node at (0, -1.2) {\tiny{$I$}};
\end{tikzpicture}
\end{gathered}
\quad&=\quad\gamma_{W,W,I}\begin{gathered}
\begin{tikzpicture}[scale=1]
\draw [ultra thick, blue, densely dotted] (0,0)  -- (0,-1) ;
\draw [ultra thick, red, -<-=.6] (0,0) -- (.87,.5) ;
\draw [ultra thick, red, -<-=.6] (0,0) -- (-.87,.5) ;
\draw (-0.17,.1)\dotsolb {right}{};
\node at (-1, 0.7) {\tiny{$W$}};
\node at (1, 0.7) {\tiny{$W$}};
\node at (0, -1.2) {\tiny{$I$}};
\end{tikzpicture}
\end{gathered}\,,\qquad& \begin{gathered}
\begin{tikzpicture}[scale=1]
\draw [ultra thick, blue, densely dotted] (0,0)  -- (0,-1) ;
\draw [ultra thick, red, -<-=.6] (0,0) -- (.87,.5) ;
\draw [ultra thick, red, -<-=.6] (0,0) -- (-.87,.5) ;
\draw (0,0)\dotsolb {right}{};
\node at (-1, 0.7) {\tiny{$W$}};
\node at (1, 0.7) {\tiny{$W$}};
\node at (0, -1.2) {\tiny{$I$}};
\end{tikzpicture}
\end{gathered}
\quad&=\quad
\gamma_{W,W,I}^{-1}\begin{gathered}
\begin{tikzpicture}[scale=1]
\draw [ultra thick, blue, densely dotted] (0,0)  -- (0,-1) ;
\draw [ultra thick, red, -<-=.6] (0,0) -- (.87,.5) ;
\draw [ultra thick, red, -<-=.6] (0,0) -- (-.87,.5);
\draw (0.17,.1)\dotsolb {right}{};
\node at (-1, 0.7) {\tiny{$W$}};
\node at (1, 0.7) {\tiny{$W$}};
\node at (0, -1.2) {\tiny{$I$}};
\end{tikzpicture}
\end{gathered}\,,
\fe
where the black dot is the 1D Majorana fermion. The four different ways of the combination of $\gamma$ and $\kappa$ encode all $\mathbb Z_2$ $W$-line of odd $\nu$ cases, summarized below:
\begin{table}[h!]
    \centering
    \begin{tabular}{c||c|c|c}
         &  $\gamma$ & FS indicator & Spin selection rule $s_W=\frac{\nu_W}{16}\,({\rm mod\,\frac{1}{2}})$\\
         \hline
       $\nu_W=1$  &  $e^{3\pi i/4}$ & $\kappa=+1$ & $s_W=\frac{1}{16}\,({\rm mod\,\frac{1}{2}})$\\
       \hline
       $\nu_W=3$  &  $e^{\pi i/4}$ & $\kappa=-1$ & $s_W=\frac{3}{16}\,({\rm mod\,\frac{1}{2}})$\\
       \hline
       $\nu_W=5$  &  $e^{3\pi i/4}$ & $\kappa=-1$ & $s_W=\frac{5}{16}\,({\rm mod\,\frac{1}{2}})$\\
       \hline
       $\nu_W=7$  &  $e^{\pi i/4}$ & $\kappa=+1$ & $s_W=\frac{7}{16}\,({\rm mod\,\frac{1}{2}})$\\\hline
    \end{tabular}
    \caption{The classification of q-type $\mathbb{Z}_2$ TDLs}
    \label{tab:qtable}
\end{table}

\section{Super-pentagon solutions} 
\label{sec:super-pentagon}
In this section, we lay out the necessary techniques to solve the super-pentagon equations associated to the fusion ring \eqref{eqn:fusion_ring}. For more details on the difference between super-pentagon equations and pentagon equation in a bosonic system, please refer to our previous work \cite{Choi:2022jqy}. The solutions to the super-pentagon equations are classified by the parameters $\nu_Z$, $\nu_W$, and $\nu_{WZ}$, which are associated to the three independent $\mathbb{Z}_2$ subcategories spanned by the subsets of TDLs
\ie
\{I, Z\}, \qquad \{I, W\}, \qquad \{I, WZ\}\,.
\fe
The fermion parity TDL $Z$ should a non-anomalous m-type line, i.e. $\nu_Z=0$. Furthermore, when the TDL $W$ is q-type, the 1d Majorana fermion on $W$ should acquires a minus sign when passing through ther fermion parity TDL $Z$. This requirement gives a new constraint:
\ie
\nu_W+\nu_{WZ}=0,4\mod 8\,.
\fe
This constraint is also satisfied when $W$ is m-type. When the system enjoys a parity symmetry, we argue a stronger constraint 
\ie
\nu_W+\nu_{WZ}=0\mod 8\,.
\fe

To avoid cluttered notation, we introduce the numeric labels for the 4 different TDLs:
\ie\label{eqn:line_numeric_label}
1~:~I\,,\quad 2~:~Z\,,\quad 3~:~W\,,\quad 4~:~WZ\,.
\fe

\subsection{
q-type $\mathbb{Z}_2$ flavor symmetry
}\label{sec:qxm}
The fusion ring is then given by 
\begin{equation}
    W^2=(1_b+1_f)I, \qquad Z^2=1_b I, \qquad (WZ) (WZ)=(1_b+1_f)I, \qquad WZ=ZW
\end{equation}
The full set of the super-pentagon equaitons in this case admits $32$ inequivalent solutions classified by
\ie
\nu_Z = 0,\,4\,,\quad \nu_W,\, \nu_{WZ} = 1,\,3,\,5,\,7\,.
\fe

Since the TDL $Z$ representing the fermion parity symmetry, which is non-anomalous, the only non-trivial F-symbol in $Z$ should be taken as 
\ie
\mathcal{F}^{Z,Z,Z}_Z=1\,,~\text{i.e.}~\nu_Z=0\,.
\fe
This requirement reduces number of solutions to 16, determined only by $\nu_W$ and $\nu_{WZ}$.

As alluded in the previous section, the $q$-type super-pentagon solutions can be classified by two gauge invariant parameters. Accordingly, the remaining 16 solutions can be labeled by the coefficients $(\gamma_{3,3,1},\,\gamma_{4,4,1})$ and the Frobenius-Schur indicators $\kappa_3,\,\kappa_4=\pm 1$ associated with $W$ and $WZ$, where the subscripts are the numeric labels \eqref{eqn:line_numeric_label} for the TDLs.  

We further impose the condition that the 1d Majorana fermion on the $q$-type acquires a minus sign when passing through the $Z$ line, i.e.,
\ie\label{eqn:1dMF_passing_Fparity}
\begin{gathered}
\begin{tikzpicture}[scale=1.5]
\draw [line,lightgray] (0,0)  -- (0,2) -- (2,2) -- (2,0) -- (0,0) ;
\draw [ultra thick,red] (1,0) -- (1,2) ;
\draw [ultra thick,blue] 
(0,1) -- (2,1) ;
\draw (1,0.8)\dotsolb {right}{}; 
\node at (0.3,0.85) {\footnotesize{$Z$}};
\node at (1.7,0.85) { \footnotesize{$Z$}};
\node at (1.45,1.7) {\footnotesize{ $W(WZ)$}};
\node at (1.45,0.3) {\footnotesize{ $W(WZ)$}};
\end{tikzpicture}
\end{gathered} \quad=\quad (-1)\quad\times\quad 
\begin{gathered}
\begin{tikzpicture}[scale=1.5]
\draw [line,lightgray] (0,0)  -- (0,2) -- (2,2) -- (2,0) -- (0,0) ;
\draw [ultra thick,red] (1,0) -- (1,2) ;
\draw [ultra thick,blue] 
(0,1) -- (2,1) ;
\draw (1,1.2)\dotsolb {right}{}; 
\node at (0.3,0.85) {\footnotesize{$Z$}};
\node at (1.7,0.85) { \footnotesize{$Z$}};
\node at (1.45,1.7) {\footnotesize{ $W(WZ)$}};
\node at (1.45,0.3) {\footnotesize{ $W(WZ)$}};
\end{tikzpicture}
\end{gathered}
\fe
where we used the numeric notation \eqref{eqn:line_numeric_label} for TDLs. A detailed analysis of this condition is given in Appendix~\ref{appen:gauging}, where we shows that it cuts down the number of solutions by a half. The remaining 8 solutions have
\ie
\left(\gamma_{3,3,1},\, \gamma_{4,4,1}\right)=\left(e^{\frac{\pi i}{4}},\, e^{\frac{3\pi i}{4}}\right)\,, \ \ {\rm or}\ \ \left(e^{\frac{3\pi i}{4}},\, e^{\frac{\pi i}{4}}\right)\,,
\fe
with more details presented in Table~\ref{tab:qtype}. The F-symbols associated to two $q$ type $W$ and $WZ$ can be found in \ref{app:cq0}, which is fully determined by the set of parameters $\gamma, \kappa$. From Table~\ref{tab:qtype}, we see that the parameters $\nu_W$ and $\nu_{WZ}$ satisfy the condition
\ie\label{eqn:WWZ_ssr}
16(s_W+s_{WZ})=\nu_W+\nu_{WZ}=0,4\mod 8 \,.
\fe


\begin{table}[h!]
\label{table:cqcm}
\centering
\footnotesize
\begin{tabular}{c||c|c|c|c}
\hline
$(\gamma_{3,3,1},\kappa_3;\gamma_{4,4,1},\kappa_4)$& $(e^{\pi i/4},1;e^{3\pi i/4},1)$  & $(e^{\pi i/4},1;e^{3\pi i/4},-1)$ & $(e^{\pi i/4},-1;e^{3\pi i/4},1)$ & $(e^{\pi i/4},-1;e^{3\pi i/4},-1)$  \\
\hline
$s_W$ & $\frac{7}{16}$ & $\frac{7}{16}$ & $\frac{3}{16}$ & $\frac{3}{16}$ \\
\hline
$s_{WZ}$ & $\frac{1}{16}$ & $\frac{5}{16}$ & $\frac{1}{16}$ & $\frac{5}{16}$ \\
\hline
$(\gamma_{3,3,1},\kappa_3;\gamma_{4,4,1},\kappa_4)$& $(e^{3\pi i/4},1;e^{\pi i/4},1)$ & $(e^{3\pi i/4},1;e^{\pi i/4},-1)$ & $(e^{3\pi i/4},-1;e^{\pi i/4},1)$ & $(e^{3\pi i/4},-1;e^{\pi i/4},-1)$ \\
\hline
$s_W$ &$\frac{1}{16}$ &$\frac{1}{16}$ &$\frac{5}{16}$ &$\frac{5}{16}$ \\
\hline
$s_{WZ}$ &$\frac{7}{16}$ &$\frac{3}{16}$ &$\frac{7}{16}$ &$\frac{3}{16}$ \\
\hline
\end{tabular}
\caption{The spin selection rules for the two q-type lines: $W$ and $WZ$.  }
\label{tab:qtype}
\end{table}


\subsection{m-type $\mathbb{Z}_2$ flavor symmetry}
\label{sec:hatcmcm}

Let us first focus on the case with $\nu_W = 2$, $4$, i.e., the fusion ring: 
\begin{equation}
    W^2=1_fI, \qquad Z^2=1_b I, \qquad (WZ) (WZ)=1_fI, \qquad WZ=ZW
\end{equation} 
The full set of the super-pentagon equations in this case admits 8 inequivalent solutions, classifying by
\ie
\nu_Z=0,\,4,\,\quad \nu_W,\,\nu_{WZ}=2,\,6\,.
\fe
Since the fermion parity TDL $Z$ is non-anomalous, its F-symbol is $\mathcal{F}^{Z,Z,Z}_Z=1 = \exp(2\pi i \frac{\nu_Z}{8})$. This reduces the number of solutions to 4, corresponding to the four choices 
\ie
\mathcal{F}^{W,W,W}_W=\pm i=\exp\left(2\pi i \frac{\nu_W}{8}\right), \qquad \mathcal{F}^{WZ,WZ,WZ}_{WZ}=\pm i=\exp\left(2\pi i \frac{\nu_{WZ}}{8}\right)\,.
\fe
We see that the parameters $\nu_W$ and $\nu_{WZ}$ also satisfy the condition \eqref{eqn:WWZ_ssr}.

Next, we consider the case with $\nu_W=0$, $4$, i.e., the fusion ring: 
\begin{equation}
    W^2=1_bI, \qquad Z^2=1_bI, \qquad (WZ) (WZ)=1_bI, \qquad WZ=ZW
\end{equation}
There are 8 solutions, classifying by
\ie
\nu_Z,\,\nu_W,\,\nu_{WZ}=0,\,4\,.
\fe
Once again, we choose $\nu_Z=0$ and there are remaining 4 solutions to the super-pentagon equations with the non-trivial F-symbols,
\ie
\mathcal{F}^{W,W,W}_W=\pm 1=\exp\left(2\pi i \frac{\nu_{WZ}}{8}\right), \qquad \mathcal{F}^{WZ,WZ,WZ}_{WZ}=\pm 1=\exp\left(2\pi i \frac{\nu_{WZ}}{8}\right)\,.
\fe
We see that the parameters $\nu_W$ and $\nu_{WZ}$ also satisfy the condition \eqref{eqn:WWZ_ssr}.

\subsection{Parity action on super-fusion categories}
\label{sec: parity}
For a given 2D CFT, one can consider a parity transformation to reverse the spatial direction:
\ie
\mathcal P:\quad x\longrightarrow -x\,,
\fe
which thus swaps the Virasoro generators $L_0$ and $\bar {L}_0$. Therefore, the spin of a state under the action of $\mathcal P$ will be reversed accordingly. In a parity-invariant CFT, on the level of TDLs, starting with defect Hilbert space $\mathcal H_{\mathcal L}$ defined by a TDL $\mathcal L$, $\mathcal P$ will induce a unitary operator $U(\mathcal P)$ mapping $\mathcal H_{\mathcal L}$ to $\mathcal H_{\mathcal L'}$ defined by $\mathcal L'$, the parity dual object of $\mathcal L$,
\ie
U(\mathcal P):\quad
\begin{gathered}
\begin{tikzpicture}[scale=1]
\path (0, 2.8);
\draw [line,lightgray] (0,0)  -- (0,2) -- (2,2) -- (2,0) -- (0,0) ;
\draw [ultra thick,red] (1,0) -- (1,2) ;
\node at (1.3, 1) {\footnotesize{$\mathcal L$}};
\node at (1, -.3) {$\mathcal H_{\mathcal L}$};
\end{tikzpicture}
\end{gathered}
\quad\longrightarrow\quad
\begin{gathered}
\begin{tikzpicture}[scale=1]
\path (0, 2.8);
\draw [line,lightgray] (0,0)  -- (0,2) -- (2,2) -- (2,0) -- (0,0) ;
\draw [ultra thick,red] (1,0) -- (1,2) ;
\node at (1.3, 1) {\footnotesize{$\mathcal L'$}};
\node at (1, -.3) {$\mathcal H_{\mathcal L'}$};
\end{tikzpicture}
\end{gathered}
\fe
Therefore the spin selection rules determined by $\mathcal L$ and $\mathcal L'$ in $\mathcal H_{\mathcal L}$ and $\mathcal H_{\mathcal L'}$ need to satisfy
\ie
s\left(|\mathcal O_{\mathcal L}\rangle\right)+s\left(U(\mathcal P)|\mathcal O_{\mathcal L}\rangle\right)\equiv 0\left(\!\!\!\!\!\!\mod\frac{1}{2}\right)\,,
\fe
where $|\mathcal O_{\mathcal L}\rangle$ denote a state in $\mathcal H_{\mathcal L}$, and $U(\mathcal P)|\mathcal O_{\mathcal L}\rangle$ is a state in $\mathcal H_{\mathcal L'}$. Clearly, for many bosonic rational CFTs, e.g. the minimal models, the TDL $\mathcal L$'s are self-dual, and thus $U(\mathcal P)$ is a non-trivial $\mathbb Z_2$ action on $\mathcal H_{\mathcal L}$ itself mapping a state of spin $s$ to one of $-s$. A typical example is the Ising fusion category, which contain a non-invertible Krammer-Wannier duality line $\mathcal N$, whose spin selection rule has been determined in \cite{Chang:2018iay} as
\ie
s\left(\mathcal H_{\mathcal N}\right)=\left\{\pm\frac{1}{16}\right\}\,.
\fe

However, in a fermionic CFT $\mathfrak T_f$, the TDLs in the super-fusion category of $\mathfrak T_f$ are generically not self-dual. A simple example is the fermionization of Ising model, which contain four TDLs,
\ie
\left\{1,\, (-1)^F,\, (-1)^{F_L}, \, (-1)^{F_R}\right\}
\,.
\fe
The parity symmetry $\mathcal P$ straightforwardly maps $(-1)^{F_L}$ to $(-1)^{F_R}$ and vice verse. In the latter section, one will clearly see that the spin selection rules of  $(-1)^{F_L}$ and $(-1)^{F_R}$ are $+\frac{1}{16}$ and $-\frac{1}{16}$ respectively, inherited from that of $\mathcal N$-line in its bosonized Ising cousin.

Based on the argument above, we will require that, in a parity invariant fermionic CFT, the set of TDLs in a super-fusion category has to be closed under the action of parity symmetry $\mathcal P$. This additional requirement will further rule out 4 solutions discussed in sec.\ref{sec:qxm} of the spin selection rules satisfying
\ie
\nu_W+\nu_{WZ}=4\mod 8\,,
\fe
An interesting question is to ask what (fermionic) CFTs can realize such (super-)fusion categories. A possible candidate answer would be some 2D fermionic CFTs with gravitational anomaly, bounded by 3D bulk theories, that are not invariant respect to parity symmetry. 

On the other hand, the left four solutions satisfying the spin selection rules
\ie
\nu_W+\nu_{WZ}=0\mod 8\,,
\fe
would be the super-fusion categories of fermionic theories $\mathfrak T_f$, whose bosonization $\mathfrak T_b$ contain the Ising category. We will realize these CFTs by stacking $\nu=1,3,5,7$ copies of Majorana fermions in Sec.\ref{sec:8_Maj}.

\section{The realizations in 2D CFTs of $\nu$ copies of Majorana Fermions}
\label{sec:8_Maj}
In this section, we argue that some of superfusion category solved in the last section \ref{sec:super-pentagon} can be realized in $\nu$ copies of 2D Majorana fermions. The additional $\mathbb{Z}_2$ symmetry is realized by $W\equiv(-1)^{F_L}$  
\begin{equation}
  (-1)^{F_L}=(-1)^{F_L^{1}} \cdots (-1)^{F_L^{\nu}}  
\end{equation}
where $(-1)^{F_L^{i}}$ acts as left fermion parity on the $i$-th copy of Majorana fermions. For more details on the 2d Majorana fermions and its bosonization, we refer to \ref{appen:2dMF} and references therein. The $\mathbb Z_8$ classification of $W$ can be seen from the defect Hilbert space $\mathcal H_{W}$ in this theory. To this end, note that the partition function of this $v$ copies of 2D Majorana reads
\begin{align}
    \mathcal Z=|\chi_0+\chi_{1/2}|^{2\nu}\,.
\end{align}
Inserting the $W$- and $WZ$-line along the time direction, the defect partition function is spelled as
\begin{align}
    \mathcal Z_{W}=2^{\nu/2}(\bar\chi_0+\bar\chi_{1/2})^\nu\chi_{1/16}^\nu\,.
\label{eq:defect_pf}
\end{align}
Therefore the defect Hilbert space $\mathcal H_{W}$ contains states of spin $\frac{\nu}{16}+\frac{1}{2}\mathbb Z$. Clearly, the spin selection rules admit a $\mathbb Z_8$-periodicity. This is known that in a 2D fermionic system, a $\mathbb{Z}_2$ symmetry admits a $\mathbb{Z}_8$ classification \cite{Kapustin:2014dxa, Karch:2019lnn}, resonating with the spin cobordism group $\Omega_2^{\textbf{spin}}(B\mathbb{Z}_2)=\mathbb{Z}_8$. Similiarly, we solve the defect parition function of $WZ$, which reads
\begin{align}
    \mathcal Z_{W}=2^{\nu/2}(\chi_0+\chi_{1/2})^\nu\bar\chi_{1/16}^\nu\,.
\label{eq:defect_pfWZ}
\end{align}
and the defect Hilbert space $\mathcal H_{WZ}$ contains states of spin $-\frac{\nu}{16}+\frac{1}{2}\mathbb Z$.

By matching with their spin selction rule, in particular, we can tell that the $W$ and $WZ$ for $\nu=2\mathbb{Z}+1$ cases is a $q$-type object. The associated super-fusion category can be found four solutions as discussed before and in ref.\cite{Chang:2022hud}. On the other hand, for $\nu=2\mathbb{Z}$, the $W$-line is of m-type. Among them, the $\nu=0$ and $\nu=4$ cases correspond to the ordinary non-anomalous and anomalous $\mathbb Z_2$-symmetry. Meanwhile, $\nu=2$ and $\nu=6$ are $\mathbb Z_2$-symmetry where the 3-way junction of $V^{W,W}_{1}$ is a fermionic type,
\ie
\begin{gathered}
\begin{tikzpicture}[scale=1]
\draw [ultra thick, densely dotted, blue] (0,0)  -- (0,-1) ;
\draw [ultra thick, red, -<-=.6] (0,0) -- (.87,.5) ;
\draw [ultra thick, red, -<-=.6] (0,0) -- (-.87,.5) ;
\draw (0,0)\dotsolb {right}{};
\node at (-1, 0.7) {\footnotesize{$W$}};
\node at (1, 0.7) {\footnotesize{$W$}};
\node at (0, -1.3) {\footnotesize{$1$}};
\end{tikzpicture}\,.
\end{gathered}
\fe
Here the black dot in the junction denotes a fermionic zero mode, and solid and dashed black lines for the $\mathbb Z_2$-line and identity respectively. 

The non-trivial F-symbols for the four $m$-type $\mathbb Z_2$-symmetry is given below:
\ie
\begin{gathered}
\begin{tikzpicture}[scale=.75]
\draw [ultra thick, red, -<-=.6] (-1,1) -- (-2,2);
\draw [ultra thick, red, -<-=.6] (-1,1) --(0,2);
\draw [ultra thick, red, -<-=.6] (0,0) --(2,2);
\draw [ultra thick, red, ->-=.6] (0,0)  -- (0,-1);
\draw [ultra thick, densely dotted, blue] (0,0) -- (-1,1) ;
\draw (-1,1)\dotsolb {}{};
\end{tikzpicture}
\end{gathered}
\quad=\quad \left(e^{-\frac{\pi i}{4}}\right)^\nu
\begin{gathered}
\begin{tikzpicture}[scale=.75]
\draw [ultra thick, red, ->-=.6] (-2,2) -- (0,0);
\draw [ultra thick, red, ->-=.6] (0,2) --(1,1);
\draw [ultra thick, red, -<-=.6] (1,1) --(2,2);
\draw [ultra thick, densely dotted, blue] (0,0)  -- (1,1);
\draw [ultra thick, red, ->-=.6] (0,0) -- (0,-1) ;
\draw (1,1)\dotsolb {}{};
\end{tikzpicture}
\end{gathered}\,,
\label{eq:Z_2_nu}
\fe
i.e. $\mathcal F^{W,W,W}_{W}=e^{-\frac{\pi i\nu}{4}}, \qquad \nu=(0,2,4,6)$.

\subsection{$\mathbb Z_2^{\nu_W=1,3,5,7}\times\mathbb Z_2^{\nu_Z=0}$  }
Based on the preceding analysis, we argue that $W$ and $WZ$ are q-type topological defect lines (TDLs) in theories with $\nu=2\mathbb{Z}+1$. Consequently, the superfusion category for these models is described by $\mathbb Z_2^{\nu_W=1,3,5,7}\times\mathbb Z_2^{\nu_Z=0}$, where the parameters $\gamma, \kappa$ for the $W, WZ$ lines are chosen as follows:
\begin{table}[h!]
\centering
\footnotesize
\begin{tabular}{|c||c|c|}
\hline
 $(\gamma_{3,3,1},\kappa_3;\gamma_{4,4,1},\kappa_4)$& $(e^{\pi i/4},1;e^{3\pi i/4},1)$  &  $(e^{\pi i/4},-1;e^{3\pi i/4},-1)$  \\
\hline
$W$ & $\frac{7}{16}$ & $\frac{3}{16}$ \\
\hline
$WZ$ & $\frac{1}{16}$ &$\frac{5}{16}$ \\
\hline
$(\gamma_{3,3,1},\kappa_3;\gamma_{4,4,1},\kappa_4)$& $(e^{3\pi i/4},1;e^{\pi i/4},1)$ &  $(e^{3\pi i/4},-1;e^{\pi i/4},-1)$ \\
\hline
$W$ &$\frac{1}{16}$ &$\frac{5}{16}$ \\
\hline
$WZ$ &$\frac{7}{16}$ &$\frac{3}{16}$ \\
\hline
\end{tabular}
\caption{Spin selection rules}
\label{tab:qtype}
\end{table}
These theories are invariant under parity symmetry; thus, following the arguments in \ref{sec: parity}, we indeed have
\ie
\nu_W+\nu_{WZ}=0\mod 8\,,
\fe

\subsection{$\mathbb Z_2^{\nu_W=2,6}\times\mathbb Z_2^{\nu_Z=0}$  }

To see why the $\nu=2, 6$ $m$-type $\mathbb Z_2$-line need a fermionic type junction, let's focus on $\nu=2$, i.e. two massless Majorana fermions case. As the two Majorana fermions compose a Dirac fermion, the 2D CFT describing it has central charge $c=1$. Its bosonization is the compact boson at the circle of radius $R=2$, i.e. the $U(1)_4$ WZW model. The CFT has four primaries with conformal weights $h=\{0,\frac{1}{8},\frac{1}{2},\frac{1}{8}^c\}$, where $h=\frac{1}{8}^c$ denotes a primary of conformal weight $\frac{1}{8}$ and $\mathbb Z_4$-conjugate to the primary of $h=\frac{1}{8}$. The four primaries amount to four Verlinde lines $\mathcal L_h$, furnishing an anomalous $\mathbb Z_4^{\omega=2}$-symmetry, where $\omega\in H^3(\mathbb Z_4,U(1))\simeq \mathbb Z_4$ captures the anomaly of $\mathbb Z_4$ in $U(1)_4$. The bosonic partition function is given by
\begin{align}
    \mathcal Z_{U(1)_4}=\left|\chi^{U(1)_4}_0\right|^2+\left|\chi^{U(1)_4}_{1/8}\right|^2+\left|\chi^{U(1)_4}_{1/2}\right|^2+\left|\chi^{U(1)_4}_{1/8^c}\right|^2\,,
\end{align}
where $\chi^{U(1)_4}_h$ denotes the character of the $U(1)_4$ primary with conformal weight $h$. Although the $\mathbb Z_4$ is anomalous, the subgroup $\mathbb Z_2\subset\mathbb Z_4$ generated by $\mathcal L_{1/2}$ is non-anomalous. One can use it to fermionize the bosonic $U(1)_4$ back to the Dirac fermion theory,
\begin{align}
    \mathcal Z_{\rm Dirac}=\left|\chi^{U(1)_4}_0+\chi^{U(1)_4}_{1/2}\right|^2\,.
\end{align}
One can honestly check that 
\begin{align}
    \chi^{U(1)_4}_0+\chi^{U(1)_4}_{1/2}=(\chi_0+\chi_{1/2})^2\,,\quad{\rm and\ thus}\quad \mathcal Z_{\rm Dirac}=\mathcal Z_{\nu=2}\,.
\end{align}
The $(-1)^{F_L}$-line, i.e. $W$, can be obtained from the line $\mathcal L_{1/8}\in\mathbb Z_4$ in its bosonic cousin via the fermionic condensation. Notice that in $U(1)_4$, we have the fusion rule
\begin{align}
    \mathcal L_{1/8}\cdot\mathcal L_{1/8}=\mathcal L_{1/2}\,,
\end{align}
While fermionization condenses the $\mathcal L_{1/2}$-line to vacuum, the $\mathcal L_{1/8}$ condenses to $W$, accompanied by a fermionic junction that encodes the condensed line $\mathcal L_{1/2}$ \cite{Zhou:2021ulc}, as depicted by fig. \ref{eq:nu=2&6}.
\ie
\begin{gathered}
\begin{tikzpicture}[scale=1]
\draw [ultra thick, black, ->-=.6] (0,0)  -- (0,-1) ;
\draw [ultra thick, red, -<-=.6] (0,0) -- (.87,.5) ;
\draw [ultra thick, red, -<-=.6] (0,0) -- (-.87,.5) ;
\node at (-1, 0.7) {\tiny{$\mathcal L_{1/8}$}};
\node at (1, 0.7) {\tiny{$\mathcal L_{1/8}$}};
\node at (0, -1.2) {\tiny{$\mathcal L_{1/2}$}};
\end{tikzpicture}
\end{gathered}=
\begin{gathered}
\begin{tikzpicture}[scale=1]
\draw [ultra thick, black] (0,0)  -- (0,-0.4) ;
\draw [ultra thick, black, ->-=.6] (0,-0.4)  -- (0.8,0.08) ;
\draw [ultra thick, densely dotted, blue] (0,-0.4)  -- (0,-1) ;
\draw [ultra thick, red, -<-=.6] (0,0) -- (.87,.5) ;
\draw [ultra thick, red, -<-=.6] (0,0) -- (-.87,.5) ;
\node at (-1, 0.7) {\tiny{$\mathcal L_{1/8}$}};
\node at (1, 0.7) {\tiny{$\mathcal L_{1/8}$}};
\node at (-0.5, -0.2) {\tiny{$\mathcal L_{1/2}$}};
\node at (1, -0.2) {\tiny{$\mathcal L_{1/2}$}};
\node at (0, -1.2) {\tiny{$1$}};
\end{tikzpicture}
\end{gathered}
\quad\xrightarrow{\rm fermionic\ condensation\ of\ \mathcal L_{1/2}}\quad
\begin{gathered}
\begin{tikzpicture}[scale=1]
\draw [ultra thick, densely dotted, blue] (0,0)  -- (0,-1) ;
\draw [ultra thick, red, -<-=.6] (0,0) -- (.87,.5) ;
\draw [ultra thick, red, -<-=.6] (0,0) -- (-.87,.5) ;
\draw (0,0)\dotsolb {right}{};
\node at (-1, 0.7) {\tiny{$W$}};
\node at (1, 0.7) {\tiny{$W$}};
\node at (0, -1.2) {\tiny{$1$}};
\end{tikzpicture}
\end{gathered}
\label{eq:nu=2&6}
\fe

Therefore, the corresponding superfusion category realized by these two models is $\mathbb Z_2^{\nu_W=2,6}\times\mathbb Z_2^{\nu_Z=0}$ . As alluded in \ref{sec:hatcmcm}, four out of eight solutions have non-anomalous $(-1)^F$ symmetry, i.e.
\ie
\mathcal F^{Z,Z,Z}_{Z}=1\,.
\fe
To find the correct one among the four solutions, we need to understand the spin selection rules of the defect Hilbert space in presence of the lines $W$ and $WZ$. A simple way to determine them is to compare with the defect partition function \eqref{eq:defect_pf}, or the defect partition function of the Dirac fermion. The latter can be also easily computed as
\ie
&\mathcal Z_{{\rm Dirac},W}=\left(\bar\chi^{U(1)_4}_0+\bar\chi^{U(1)_4}_{1/2}\right)\left(\chi^{U(1)_4}_{1/8}+\chi^{U(1)_4}_{1/8^c}\right)\,,\notag\\
&\mathcal Z_{{\rm Dirac},WZ}=\left(\bar\chi^{U(1)_4}_{1/8}+\bar\chi^{U(1)_4}_{1/8^c}\right)\left(\chi^{U(1)_4}_{0}+\chi^{U(1)_4}_{1/2}\right)
\fe
So the spin selection rules read as
\ie
&\nu=2:\quad 
s_{W}=\frac{1}{8}\left({\rm mod}\ \frac{1}{2}\right)\,,\ \quad
s_{WZ}=-\frac{1}{8}\left({\rm mod}\ \frac{1}{2}\right)\,;\notag\\
&\nu=6:\quad 
s_{W}=-\frac{1}{8}\left({\rm mod}\ \frac{1}{2}\right)\,,\ \quad
s_{WZ}=\frac{1}{8}\left({\rm mod}\ \frac{1}{2}\right)\,.
\label{eq:spin_selection}
\fe
Therefore we conclude that $W$ and $WZ$ must have different spin selection rules. If $W$ is the second element in $\mathbb Z_8^{\nu}$, $WZ$ must correspond to the sixth, or vice verse.

According to \eqref{eq:Z_2_nu} and \eqref{eq:spin_selection}, there are only two solutions out of eight satisfying the spin selection rules \eqref{eq:spin_selection} corresponding to the two cases of $\nu=2$ and $6$ respectively, i.e.
\ie
\mathcal F^{W,W,W}_{W}=\pm i\,,\qquad
\mathcal F^{WZ,WZ,WZ}_{WZ}=\mp i
\fe
In both the $\nu=2$ and $6$ cases, we have the following $F$-symbol (the numerical label stated in \eqref{eqn:line_numeric_label})
\ie
\begin{gathered}
\begin{tikzpicture}[scale=1]
\draw [line,lightgray] (0,0)  -- (0,2) -- (2,2) -- (2,0) -- (0,0) ;
\draw [ultra thick, red] (1,0) -- (1,2) ;
\draw [ultra thick, blue] (1,.7) .. controls (1,1) and (1.5,1) .. (2,1)
(0,1) .. controls (.5,1) and (1,1) .. (1,1.3) ;
\node at (0.2,1.15) {\tiny $2$};
\node at (1.8,0.85) {\tiny $2$};
\node at (1.3,1.15) {\tiny $4(3)$};
\node at (1.3,1.8) {\tiny $3(4)$};
\node at (1.3,0.2) {\tiny $3(4)$};
\end{tikzpicture}
\end{gathered}
\quad = \quad (-1)\quad\times\quad
\begin{gathered}
\begin{tikzpicture}[scale=1]
\draw [line,lightgray] (0,0)  -- (0,2) -- (2,2) -- (2,0) -- (0,0) ;
\draw [ultra thick,red] (1,0) -- (1,2) ;
\draw [ultra thick,blue] (1,1.3) .. controls (1,1) and (1.5,1) .. (2,1)
(0,1) .. controls (.5,1) and (1,1) .. (1,.7) ;
\node at (0.2,1.15) {\tiny $2$};
\node at (1.8,0.85) {\tiny $2$};
\node at (1.3,0.85) {\tiny $4(3)$};
\node at (1.3,1.8) {\tiny $3(4)$};
\node at (1.3,0.2) {\tiny $3(4)$};
\end{tikzpicture}
\end{gathered}
\,,
\label{graph:nu26_-1F}
\fe
i.e. $\mathcal F^{Z,\, W,\, Z}_{W}=-1$ for $\nu=2$ and $6$, where the red lines denote for $W$, and the blue line for $Z$.

\subsection{ $\mathbb Z_2^{\nu_W=0,4}\times\mathbb Z_2^{\nu_Z=0}$  }
For $\nu=0,4$, the Spin selection rules can be also read off from eq.\ref{eq:defect_pf}, and it shows that
\ie
&\nu=0:\quad 
s_{W}=0\left({\rm mod}\ \frac{1}{2}\right)\,,\ \quad
s_{WZ}=0\left({\rm mod}\ \frac{1}{2}\right)\,;\notag\\
&\nu=4:\quad 
s_{W}=\frac{1}{4}\left({\rm mod}\ \frac{1}{2}\right)\,,\ \quad
s_{WZ}=\frac{1}{4}\left({\rm mod}\ \frac{1}{2}\right)\,.
\label{eq:spin_selection}
\fe
It implies that, for $\nu=0$, both $W$ and $WZ$ are anomaly free, while for $\nu=4$, both of them are anomalous. By searching out the pentagon solutions, we can find uniquely one solution for each case. For both $\nu=0,4$, the mixed 't Hooft anomaly between $W$ and $Z$ is \emph{trivial}, i.e.
\ie
\begin{gathered}
\begin{tikzpicture}[scale=1]
\draw [line,lightgray] (0,0)  -- (0,2) -- (2,2) -- (2,0) -- (0,0) ;
\draw [ultra thick, red] (1,0) -- (1,2) ;
\draw [ultra thick, blue] (1,.7) .. controls (1,1) and (1.5,1) .. (2,1)
(0,1) .. controls (.5,1) and (1,1) .. (1,1.3) ;
\node at (0.2,1.15) {\tiny $2$};
\node at (1.8,0.85) {\tiny $2$};
\node at (1.3,1.15) {\tiny $4(3)$};
\node at (1.3,1.8) {\tiny $3(4)$};
\node at (1.3,0.2) {\tiny $3(4)$};
\end{tikzpicture}
\end{gathered}
\quad = \quad (+1)\quad\times\quad
\begin{gathered}
\begin{tikzpicture}[scale=1]
\draw [line,lightgray] (0,0)  -- (0,2) -- (2,2) -- (2,0) -- (0,0) ;
\draw [ultra thick,red] (1,0) -- (1,2) ;
\draw [ultra thick,blue] (1,1.3) .. controls (1,1) and (1.5,1) .. (2,1)
(0,1) .. controls (.5,1) and (1,1) .. (1,.7) ;
\node at (0.2,1.15) {\tiny $2$};
\node at (1.8,0.85) {\tiny $2$};
\node at (1.3,0.85) {\tiny $4(3)$};
\node at (1.3,1.8) {\tiny $3(4)$};
\node at (1.3,0.2) {\tiny $3(4)$};
\end{tikzpicture}
\end{gathered}
\,,
\label{graph:nu04_-1F}
\fe

\section{Application}
\label{sec:app}
So far, we have solved the super-pentagon equations associated to the fusion ring \eqref{eqn:fusion_ring} and discussed (parts of) their realizations in fermionic CFTs. We now apply these results to specific examples. It is important to notice that, although our discussion is focused on the TDL in fermionic CFTs, the (super-)pentagon solutions, i.e. the 't Hooft anomalies, are \emph{rigid} against deformation from CFTs to gapped phases. Therefore we would like to apply these results to some fermionic gapped models obtained from relevant deformations of CFTs. 

\subsection{$\mathcal N=2$ minimal model $A_2$ with the least relevant deformation}
\label{sec:N=2_rel_def}
The $\mathcal N=2$ minimal model $A_2$ admits a LG formalism given by a homogeneous superpotential
\ie
\mathcal W(\Phi)=\frac{1}{3}\Phi^3
\fe
It has central charge $c=1$, and its bosonization is a compact boson at radius $R=\sqrt{12}$, i.e. $U(1)_{12}$. Therefore, in the bosonic theory, there are $12$ Verlinde lines corresponding to an anomalous $\mathbb Z^{\omega=6}_{12}$ symmetry \cite{Chang:2018iay}, where $\omega=6$ capture its `t Hooft anomaly $\omega\in H^3(\mathbb Z_{12}, U(1))\simeq {\mathbb Z_{12}}$. The associated $12$ primaries $\phi_{k}$ have thus conformal weight 
\ie
h_k=\frac{k^2}{24}\,,\quad {\rm for}\quad k=0,1,\cdots, 11\,.
\fe
The $\mathcal N=2$ $A_2$ CFT is obtained by fermionization with respect to $\mathcal L_{6}$, the generator of $\mathbb Z_2$ subgroup of $\mathbb Z_{12}$. After fermionization, the fermionic theory has a super-fusion category
\ie
\mathfrak C_{A_2}=\left(\mathbb Z_3\times \mathbb Z_2^{\nu_W=2}\right)\times \mathbb Z_2^{\nu_Z=0}\,,
\fe
Here $\mathbb Z_2^{\nu_Z=0}$, referring to $(-1)^F$, is an emergent quantum symmetry, $\mathbb Z_3$ is a non-anomalous bosonic symmetry, rotating $\Phi\rightarrow e^{\frac{2\pi i}{3}}\Phi$, $\mathbb Z_2^{\nu_W=2}$ is the $\mathbb Z_2$ subgroup of the $U(1)_R$ symmetry. 

Now we add a relevant deformation,
\ie
\mathcal W(\Phi)=\frac{1}{3}\Phi^3-\lambda \Phi\,.
\fe
Integrating out the Grassmannian variables, we have the potential in real space
\ie
V(\phi, \psi)=|\phi^2-\lambda|^2-\lambda\phi\psi^2+h.c\,,
\fe
where $\psi^2=\psi_L\psi_R$. Clearly, one can see that the $\mathbb Z_3$ rotation is broken, but the symmetries
\ie
W:\quad \phi\rightarrow -\phi\,,\quad \psi_L\rightarrow -\psi_L\,,\ \quad \psi_R\rightarrow \psi_R\,,\notag\\
Z:=(-1)^F:\quad \phi\rightarrow \phi\,,\quad \psi_L\rightarrow -\psi_L\,,\ \quad \psi_R\rightarrow -\psi_R\,.
\fe
are still preserved. Since the $\mathcal N=2$ chiral multiplet $\Phi$ contains a Dirac fermion, the $W$-symmetry corresponds $\nu=2$ case in the $\mathbb Z_8$-classification.

However now, the deformed model is gapped, having two vacua
\ie
|0\rangle=\left|\sqrt{\lambda}\right\rangle\,,\quad {\rm and}\ \quad |1\rangle=\left|-\sqrt{\lambda}\right\rangle\,,
\fe
determined by the $F$-term vacua equation
\ie
\phi^2=\lambda\,.
\fe
Therefore, the $W$-symmetry is spontaneously broken, and flips the two vacua
\ie
W\left|0\right\rangle=\left|1\right\rangle\,.
\fe
Therefore, we can consider a soliton configuration, schematically described by 
\ie
|{\rm sol}\rangle\,=\,
\begin{gathered}
\begin{tikzpicture}[scale=1]
            \draw [thick, line] (0,-.1) -- (0,1.6);
            \draw [densely dashed, ultra thick, black, ->-=.6] (.2, 0) -- (1.2, 0);
            \draw [densely dashed, ultra thick, ->-=.7] (1.2, 0) -- (2.2, 0);
            \draw [red, ultra thick, -<-=.60] (1.2, 0) -- (1.2, 1.5);
            \draw [thick, line] (2.4,-.1) -- (2.6, .85) -- (2.4, 1.6);
 \filldraw[black] (1.2,0) circle (2pt);
 \node at (1.5, 1.4) {\footnotesize $W$};
 \node at (.35, .3) {\footnotesize $|0\rangle$};
 \node at (2.1, .3) {\footnotesize $|1\rangle$};
\end{tikzpicture}
\end{gathered}
\,,
\fe
where the dashed line denotes the vacua, and red line for $W$, whose symmetry is now broken. A $(-1)^F$-line acting on the state, after resolving its 4-way junction with $\eta$, can thus be drawn as
\ie
(-1)^F|{\rm sol}\rangle\,=\,
\begin{gathered}
\begin{tikzpicture}[scale=1]
            \draw [thick, line] (0,-.1) -- (0,1.6);
            \draw [ultra thick, densely dashed, ->-=.6] (.2, 0) -- (1.2, 0);
            \draw [ultra thick, densely dashed, ->-=.7] (1.2, 0) -- (2.2, 0);
            \draw [ultra thick, red, -<-=.60] (1.2, 0) -- (1.2, 0.75);
            \draw [ultra thick, red, -<-=.60] (1.2, .75) -- (1.2, 1.5);
            \draw [thick, line] (2.4,-.1) -- (2.6, .85) -- (2.4, 1.6);
 \filldraw[black] (1.2,0) circle (2pt);
 \node at (1.5, 1.4) {\footnotesize $W$};
 \node at (.5, 1.1) {\footnotesize $(-1)^F$};
 \node at (.35, .3) {\footnotesize $|0\rangle$};
 \node at (2.1, .3) {\footnotesize $|1\rangle$};
 \draw [ ultra thick, blue, ->-=.6] (.2, 0.75) .. controls ( .7,.75) and (1.2,.75) .. (1.2, 0.6);
\draw [ ultra thick, blue, ->-=.6] (1.2, 0.9) .. controls (1.2, .75) and (1.8, .75) .. (2.3, 0.75);
\end{tikzpicture}
\end{gathered}
\fe
However, different ways of resolving the 4-way junction between the lines of $(-1)^F$ and $W$ is ambiguous and up to a sign, which actually measures the 't Hooft anomaly between them, as discussed in \eqref{graph:nu26_-1F}. Therefore, $(-1)^F$ has to act on the solitonic state projectively. To see it, we successively act $(-1)^F$ twice on $|{\rm sol}\rangle$, and have
\ie
(-1)^{2F}|{\rm sol}\rangle\,&=\,
\begin{gathered}
\begin{tikzpicture}[scale=1]
            \draw [thick, line] (0,-.1) -- (0,1.6);
            \draw [ultra thick, densely dashed, ->-=.6] (.2, 0) -- (1.2, 0);
            \draw [ultra thick, densely dashed, ->-=.7] (1.2, 0) -- (2.2, 0);
            \draw [ ultra thick, red] (1.2, 0) -- (1.2, 1.5);
            \draw [thick, line] (2.4,-.1) -- (2.6, .85) -- (2.4, 1.6);
 \filldraw[black] (1.2,0) circle (2pt);
 \node at (1.45, 1.45) {\footnotesize $W$};
 \node at (.5, 1.45) {\footnotesize $(-1)^F$};
 \draw [ ultra thick, blue, ->-=.6] (.2, 0.57) .. controls ( .7,0.57) and (1.2,0.57) .. (1.2, 0.42);
 \draw [ ultra thick, blue, ->-=.6] (1.2, 0.72) .. controls (1.2, 0.57) and (1.8, 0.57) .. (2.3, 0.57);
\draw [ ultra thick, blue, ->-=.6] (.2, 1.10) .. controls ( .7,1.10) and (1.2, 1.10) .. (1.2, .95);
 \draw [ultra thick, blue, ->-=.6] (1.2, 1.25) .. controls (1.2, 1.1) and (1.8, 1.1) .. (2.3, 1.1);
\end{tikzpicture}
\end{gathered}
\,=\,-\,
\begin{gathered}
\begin{tikzpicture}[scale=1]
            \draw [thick, line] (0,-.1) -- (0,1.6);
            \draw [densely dashed, ultra thick, ->-=.6] (.2, 0) -- (1.2, 0);
            \draw [ultra thick, densely dashed, ->-=.7] (1.2, 0) -- (2.2, 0);
            \draw [ ultra thick, red] (1.2, 0) -- (1.2, 1.5);
            \draw [thick, line] (2.4,-.1) -- (2.6, .85) -- (2.4, 1.6);
 \filldraw[black] (1.2,0) circle (2pt);
 \node at (1.45, 1.45) {\footnotesize $W$};
 \node at (.5, 1.45) {\footnotesize $(-1)^F$};
 \draw [ultra thick, blue, ->-=.6] (.2, 0.57) .. controls ( .7,0.57) and (1.2,0.57) .. (1.2, 0.72);
 \draw [ultra thick, blue, ->-=.6] (1.2, 0.42) .. controls (1.2, 0.57) and (1.8, 0.57) .. (2.3, 0.57);
\draw [ ultra thick, blue, ->-=.6] (.2, 1.10) .. controls ( .7,1.10) and (1.2, 1.10) .. (1.2, .95);
 \draw [ ultra thick, blue, ->-=.6] (1.2, 1.25) .. controls (1.2, 1.1) and (1.8, 1.1) .. (2.3, 1.1);
\end{tikzpicture}
\end{gathered}
\, = \, -|{\rm sol}\rangle\,.
\label{eq:anomaly_twice}
\fe
We thus conclude that the (anti-)solitonic states must take fractional fermion number\footnote{The feature of fractional fermion numbers in solitons was initially discussed in \cite{Fendley:1991ve, Fendley:1993pi}, and recently re-established by one of the present authors in \cite{Chen:2025qub}.},
\ie
    F_{\text{(anti-)sol}}=\pm\frac{1}{2}\,.
\fe

\subsection{$\mathcal N=1$ minimal model $A_2$ with the least relevant deformation}

It is well-known that in 2d $\mathcal{N}=(1,1)$ theories, there is no conserved fermion number operator $F$; only its $\mathbb{Z}_2$ reduction $(-1)^F$, the fermion parity, survives as a symmetry. Moreover, whether $(-1)^F$ itself remains well-defined in the soliton sector is a subtle dynamical question; see App.~\ref{appix:fermionnumber} for a detailed discussion. Nevertheless, in certain settings --- such as theories in finite volume --- the kink multiplet does admit a well-defined $(-1)^F$. 

In such cases, our $\mathcal N=2$ discussion can be generalized to the $\mathcal N=1$ case as a consistent check, where the deformed $A_2$ model can be described by a LG-superpotential
\ie
\mathcal W_{\mathcal N=1}(\Phi)=\frac{1}{3}\Phi^3-\lambda\Phi\,,
\label{eq:LG_N=1}
\fe
where $\Phi$ now is a real superfield. The potential in real space is given by
\ie
V(\phi, \psi)=(\phi^2-\lambda)^2-i\lambda\phi\psi_L\psi_R\,.
\fe
Once again we have two $\mathbb Z_2$-symmetries:
\ie
W:\quad \phi\rightarrow -\phi\,,\quad \psi_L\rightarrow -\psi_L\,,\ \quad \psi_R\rightarrow \psi_R\,,\notag\\
(-1)^F:\quad \phi\rightarrow \phi\,,\quad \psi_L\rightarrow -\psi_L\,,\ \quad \psi_R\rightarrow -\psi_R\,.
\fe
But now the $W$-line acting on a single Majorana fermion, and thus a $q$-type object. Now we have a $q$-type line spontaneously broken, switching the two vacua
\ie
W |0\rangle=|1\rangle\,.
\fe
However now because the $q$-type line can carry zero modes, and thus the (anti-)soliton states are at least two-folded degenerate \cite{Witten:1977xv, Schoutens:1990vb, Ahn:1990gn}, i.e.
\ie
|{\rm sol;\,\alpha}\rangle\,=\,
\begin{gathered}
\begin{tikzpicture}[scale=1]
            \draw [thick, line] (0,-.1) -- (0,1.6);
            \draw [densely dashed, ultra thick, black, ->-=.6] (.2, 0) -- (1.2, 0);
            \draw [densely dashed, ultra thick, ->-=.7] (1.2, 0) -- (2.2, 0);
            \draw [red, ultra thick, -<-=.60] (1.2, 0) -- (1.2, 1.5);
            \draw [thick, line] (2.4,-.1) -- (2.6, .85) -- (2.4, 1.6);
 \filldraw[black] (1.2,0) circle (2pt);
 \node at (1.7, -.3) {\footnotesize $\alpha =0,1$};
 \node at (1.5, 1.4) {\footnotesize $W$};
 \node at (.35, .3) {\footnotesize $|0\rangle$};
 \node at (2.1, .3) {\footnotesize $|1\rangle$};
\end{tikzpicture}
\end{gathered}
\,,
\fe
where $\alpha=0$ or $1$ denotes if there is a zero mode on the junction.

The bosonization of this $\mathcal N=1$ model \eqref{eq:LG_N=1} is the tri-critical Ising model, where there exists a Kramer-Wannier $\mathcal N$-line of spin selection rule $s_{\mathcal N}=\pm\frac{1}{16}$. Therefore, from the table \ref{tab:qtype}, we have $\kappa_W=1$. Now we will use the $F$-moves to show that, different from the $\mathcal N=2$ case, the two soliton states $|{\rm sol;\,\alpha}\rangle$ both have well-defined integer fermionic numbers.

Following the argument in Sec.\ref{sec:N=2_rel_def}, it's enough to show that there is no mixed 't Hooft anomaly between $W$ and $(-1)^F$, i.e. the 4-way junction resolved without ambiguity,
\ie
\begin{gathered}
\begin{tikzpicture}[scale=1]
            \draw [thick, line] (0,-.1) -- (0,1.6);
            \draw [ultra thick, densely dashed, ->-=.6] (.2, 0) -- (1.2, 0);
            \draw [ultra thick, densely dashed, ->-=.7] (1.2, 0) -- (2.2, 0);
            \draw [ultra thick, red, -<-=.60] (1.2, 0) -- (1.2, 0.75);
            \draw [ultra thick, red, -<-=.60] (1.2, .75) -- (1.2, 1.5);
            \draw [thick, line] (2.4,-.1) -- (2.6, .85) -- (2.4, 1.6);
 \filldraw[black] (1.2,0) circle (2pt);
 \node at (1.5, 1.4) {\footnotesize $W$};
 \node at (.5, 1.1) {\footnotesize $(-1)^F$};
 \node at (.35, .3) {\footnotesize $|0\rangle$};
 \node at (2.1, .3) {\footnotesize $|1\rangle$};
 \draw [ ultra thick, blue, ->-=.6] (.2, 0.75) .. controls ( .7,.75) and (1.2,.75) .. (1.2, 0.6);
\draw [ ultra thick, blue, ->-=.6] (1.2, 0.9) .. controls (1.2, .75) and (1.8, .75) .. (2.3, 0.75);
\node at (1.2, -.3) {\footnotesize $\alpha$};
\end{tikzpicture}
\end{gathered}
\,=\,
\begin{gathered}
\begin{tikzpicture}[scale=1]
            \draw [thick, line] (0,-.1) -- (0,1.6);
            \draw [ultra thick, densely dashed, ->-=.6] (.2, 0) -- (1.2, 0);
            \draw [ultra thick, densely dashed, ->-=.7] (1.2, 0) -- (2.2, 0);
            \draw [ultra thick, red, -<-=.60] (1.2, 0) -- (1.2, 0.75);
            \draw [ultra thick, red, -<-=.60] (1.2, .75) -- (1.2, 1.5);
            \draw [thick, line] (2.4,-.1) -- (2.6, .85) -- (2.4, 1.6);
 \filldraw[black] (1.2,0) circle (2pt);
 \node at (1.5, 1.4) {\footnotesize $W$};
 \node at (.5, 1.1) {\footnotesize $(-1)^F$};
 \node at (.35, .3) {\footnotesize $|0\rangle$};
 \node at (2.1, .3) {\footnotesize $|1\rangle$};
 \draw [ ultra thick, blue, ->-=.6] (.2, 0.75) .. controls ( .7,.75) and (1.2,.75) .. (1.2, 0.9);
\draw [ ultra thick, blue, ->-=.6] (1.2, 0.6) .. controls (1.2, .75) and (1.8, .75) .. (2.3, 0.75);
\node at (1.2, -.3) {\footnotesize $\alpha$};
\end{tikzpicture}
\end{gathered}
\label{eq:4-way_junc_N=1}
\fe

To prove this equality, we use the following $F$-moves:
\ie
&\hspace{-.5cm}\begin{pmatrix}
\begin{gathered}
\begin{tikzpicture}[scale=.3]
\draw [ultra thick, blue] (-1,1) -- (-2,2);
\draw [ultra thick, red] (-1,1) --(0,2);
\draw [ultra thick, blue] (0,0) --(2,2);
\draw [ultra thick, red] (0,0)  -- (0,-1);
\draw [ultra thick, red] (0,0) -- (-1,1) ;
\end{tikzpicture}
\end{gathered}
\\
\begin{gathered}
\begin{tikzpicture}[scale=.3]
\draw [ultra thick, blue] (-1,1) -- (-2,2);
\draw [ultra thick, red] (-1,1) --(0,2);
\draw [ultra thick, blue] (0,0) --(2,2);
\draw [ultra thick, red] (0,0)  -- (0,-1);
\draw [ultra thick, red] (0,0) -- (-1,1) ;
\draw (0,0)\dotsolb {}{};
\draw (-1,1)\dotsolb {}{};
\end{tikzpicture}
\end{gathered}
\\
\begin{gathered}
\begin{tikzpicture}[scale=.3]
\draw [ultra thick, blue] (-1,1) -- (-2,2);
\draw [ultra thick, red] (-1,1) --(0,2);
\draw [ultra thick, blue] (0,0) --(2,2);
\draw [ultra thick, red] (0,0)  -- (0,-1);
\draw [ultra thick, red] (0,0) -- (-1,1) ;
\draw (0,0)\dotsolb {}{};
\end{tikzpicture}
\end{gathered}
\\
\begin{gathered}
\begin{tikzpicture}[scale=.3]
\draw [ultra thick, blue] (-1,1) -- (-2,2);
\draw [ultra thick, red] (-1,1) --(0,2);
\draw [ultra thick, blue] (0,0) --(2,2);
\draw [ultra thick, red] (0,0)  -- (0,-1);
\draw [ultra thick, red] (0,0) -- (-1,1) ;
\draw (-1,1)\dotsolb {}{};
\end{tikzpicture}
\end{gathered}
\end{pmatrix}
=
\begin{large}
\frac{1}{2}\begin{pmatrix}
1 & i & 0 & 0 &\\
i & -1 & 0 & 0 & \\
0 & 0 & -i  & -1 &  \\
0 & 0 & 1 & -i &  \\
\end{pmatrix}
\end{large}
\cdot
\begin{pmatrix}
\begin{gathered}
\begin{tikzpicture}[scale=.3]
\draw [ultra thick, blue] (0,0) -- (-2,2);
\draw [ultra thick, red] (1,1) --(0,2);
\draw [ultra thick, blue] (1,1) --(2,2);
\draw [ultra thick, red] (0,0)  -- (0,-1);
\draw [ultra thick, red] (0,0) -- (1,1) ;
\end{tikzpicture}
\end{gathered}
\\
\begin{gathered}
\begin{tikzpicture}[scale=.3]
\draw [ultra thick, blue] (0,0) -- (-2,2);
\draw [ultra thick, red] (1,1) --(0,2);
\draw [ultra thick, blue] (1,1) --(2,2);
\draw [ultra thick, red] (0,0)  -- (0,-1);
\draw [ultra thick, red] (0,0) -- (1,1) ;
\draw (1,1)\dotsolb {}{};
\draw (0,0)\dotsolb {}{};
\end{tikzpicture}
\end{gathered}
\\
\begin{gathered}
\begin{tikzpicture}[scale=.3]
\draw [ultra thick, blue] (0,0) -- (-2,2);
\draw [ultra thick, red] (1,1) --(0,2);
\draw [ultra thick, blue] (1,1) --(2,2);
\draw [ultra thick, red] (0,0)  -- (0,-1);
\draw [ultra thick, red] (0,0) -- (1,1) ;
\draw (0,0)\dotsolb {}{};
\end{tikzpicture}
\end{gathered}
\\
\begin{gathered}
\begin{tikzpicture}[scale=.3]
\draw [ultra thick, blue] (0,0) -- (-2,2);
\draw [ultra thick, red] (1,1) --(0,2);
\draw [ultra thick, blue] (1,1) --(2,2);
\draw [ultra thick, red] (0,0)  -- (0,-1);
\draw [ultra thick, red] (0,0) -- (1,1) ;
\draw (1,1)\dotsolb {}{};
\end{tikzpicture}
\end{gathered}
\end{pmatrix}\,,
\fe
where the F-symbol is a $4\times 4$ matrix encoded in $\mathcal F^{(-1)^F,\,W,\,(-1)^F}_W\left((-1)^FW,\,(-1)^FW\right)$. Therefore, we show the following moves:
\ie
\begin{gathered}
\begin{tikzpicture}[scale=1]
            \draw [ultra thick, densely dashed, ->-=.6] (.2, 0) -- (1.2, 0);
            \draw [ultra thick, densely dashed, ->-=.7] (1.2, 0) -- (2.2, 0);
            \draw [ultra thick, red, -<-=.60] (1.2, 0) -- (1.2, 0.75);
            \draw [ultra thick, red, -<-=.60] (1.2, .75) -- (1.2, 1.5);
 \filldraw[black] (1.2,0) circle (2pt);
 \node at (1.5, 1.4) {\footnotesize $W$};
 \node at (.5, 1.1) {\footnotesize $(-1)^F$};
 \draw [ ultra thick, blue, ->-=.6] (.2, 0.75) .. controls ( .7,.75) and (1.2,.75) .. (1.2, 1);
\draw [ ultra thick, blue, ->-=.6] (1.2, 0.5) .. controls (1.2, .75) and (1.8, .75) .. (2.3, 0.75);
\node at (1.2, -.3) {\footnotesize $\alpha$};
\end{tikzpicture}
\end{gathered}
&\,=\,
\frac{1}{2}\,
\begin{gathered}
\begin{tikzpicture}[scale=1]
            \draw [ultra thick, densely dashed, ->-=.6] (.2, 0) -- (1.2, 0);
            \draw [ultra thick, densely dashed, ->-=.7] (1.2, 0) -- (2.2, 0);
            \draw [ultra thick, red, -<-=.60] (1.2, 0) -- (1.2, 0.75);
            \draw [ultra thick, red, -<-=.60] (1.2, .75) -- (1.2, 1.5);
 \filldraw[black] (1.2,0) circle (2pt);
 \node at (1.5, 1.4) {\footnotesize $W$};
 \node at (.5, 1.1) {\footnotesize $(-1)^F$};
 \draw [ ultra thick, blue, ->-=.6] (.2, 0.75) .. controls ( .7,.75) and (1.2,.75) .. (1.2, 0.5);
\draw [ ultra thick, blue, ->-=.6] (1.2, 1) .. controls (1.2, .75) and (1.8, .75) .. (2.3, 0.75);
\node at (1.2, -.3) {\footnotesize $\alpha$};
\end{tikzpicture}
\end{gathered}
\,+\,\frac{i}{2}\,
\begin{gathered}
\begin{tikzpicture}[scale=1]
            \draw [ultra thick, densely dashed, ->-=.6] (.2, 0) -- (1.2, 0);
            \draw [ultra thick, densely dashed, ->-=.7] (1.2, 0) -- (2.2, 0);
            \draw [ultra thick, red, -<-=.60] (1.2, 0) -- (1.2, 0.75);
            \draw [ultra thick, red, -<-=.60] (1.2, .75) -- (1.2, 1.5);
 \filldraw[black] (1.2,0) circle (2pt);
 \node at (1.5, 1.4) {\footnotesize $W$};
 \node at (.5, 1.1) {\footnotesize $(-1)^F$};
 \draw [ ultra thick, blue, ->-=.6] (.2, 0.75) .. controls ( .7,.75) and (1.2,.75) .. (1.2, 0.5);
\draw [ ultra thick, blue, ->-=.6] (1.2, 1) .. controls (1.2, .75) and (1.8, .75) .. (2.3, 0.75);
\node at (1.2, -.3) {\footnotesize $\alpha$};
 \filldraw[black] (1.2, 0.5) circle (1.5pt);
 \filldraw[black] (1.2, 1) circle (1.5pt);
\end{tikzpicture}
\end{gathered}
\notag\\
&\,=\,
\frac{1}{2}\,
\begin{gathered}
\begin{tikzpicture}[scale=1]
            \draw [ultra thick, densely dashed, ->-=.6] (.2, 0) -- (1.2, 0);
            \draw [ultra thick, densely dashed, ->-=.7] (1.2, 0) -- (2.2, 0);
            \draw [ultra thick, red, -<-=.60] (1.2, 0) -- (1.2, 0.75);
            \draw [ultra thick, red, -<-=.60] (1.2, .75) -- (1.2, 1.5);
 \filldraw[black] (1.2,0) circle (2pt);
 \node at (1.5, 1.4) {\footnotesize $W$};
 \node at (.5, 1.1) {\footnotesize $(-1)^F$};
 \draw [ ultra thick, blue, ->-=.6] (.2, 0.75) .. controls ( .7,.75) and (1.2,.75) .. (1.2, 0.5);
\draw [ ultra thick, blue, ->-=.6] (1.2, 1) .. controls (1.2, .75) and (1.8, .75) .. (2.3, 0.75);
\node at (1.2, -.3) {\footnotesize $\alpha$};
\end{tikzpicture}
\end{gathered}
\,+\,\frac{i}{2}\frac{\alpha_{3,2,4}}{\beta_{2,4,3}}\,
\begin{gathered}
\begin{tikzpicture}[scale=1]
            \draw [ultra thick, densely dashed, ->-=.6] (.2, 0) -- (1.2, 0);
            \draw [ultra thick, densely dashed, ->-=.7] (1.2, 0) -- (2.2, 0);
            \draw [ultra thick, red, -<-=.60] (1.2, 0) -- (1.2, 0.75);
            \draw [ultra thick, red, -<-=.60] (1.2, .75) -- (1.2, 1.5);
 \filldraw[black] (1.2,0) circle (2pt);
 \node at (1.5, 1.4) {\footnotesize $W$};
 \node at (.5, 1.1) {\footnotesize $(-1)^F$};
 \draw [ ultra thick, blue, ->-=.6] (.2, 0.75) .. controls ( .7,.75) and (1.2,.75) .. (1.2, 0.5);
\draw [ ultra thick, blue, ->-=.6] (1.2, 1) .. controls (1.2, .75) and (1.8, .75) .. (2.3, 0.75);
\node at (1.2, -.3) {\footnotesize $\alpha$};
 \filldraw[black] (1.2, 0.7) circle (1.5pt);
 \filldraw[black] (1.2, .8) circle (1.5pt);
\end{tikzpicture}
\end{gathered}
\,=\,
(+1)\,
\begin{gathered}
\begin{tikzpicture}[scale=1]
            \draw [ultra thick, densely dashed, ->-=.6] (.2, 0) -- (1.2, 0);
            \draw [ultra thick, densely dashed, ->-=.7] (1.2, 0) -- (2.2, 0);
            \draw [ultra thick, red, -<-=.60] (1.2, 0) -- (1.2, 0.75);
            \draw [ultra thick, red, -<-=.60] (1.2, .75) -- (1.2, 1.5);
 \filldraw[black] (1.2,0) circle (2pt);
 \node at (1.5, 1.4) {\footnotesize $W$};
 \node at (.5, 1.1) {\footnotesize $(-1)^F$};
 \draw [ ultra thick, blue, ->-=.6] (.2, 0.75) .. controls ( .7,.75) and (1.2,.75) .. (1.2, 0.5);
\draw [ ultra thick, blue, ->-=.6] (1.2, 1) .. controls (1.2, .75) and (1.8, .75) .. (2.3, 0.75);
\node at (1.2, -.3) {\footnotesize $\alpha$};
\end{tikzpicture}
\end{gathered}
\,,
\fe
where the lines $(-1)^F$,  $W$, and $(-1)^FW$ are labeled by $2,3,4$ respectively as before, and
\ie
\alpha_{3,2,4}=1\,,\quad
\beta_{2,4,3}=i\,.
\fe
In the derivation, we have used \eqref{eq:fermion_action} to move the zero modes away from the two junctions onto line $(-1)^FW$, and annihilate them. One can actually try different routes to prove \eqref{eq:4-way_junc_N=1}, for example,
\ie
\begin{gathered}
\begin{tikzpicture}[scale=1]
            \draw [ultra thick, densely dashed, ->-=.6] (.2, 0) -- (1.2, 0);
            \draw [ultra thick, densely dashed, ->-=.7] (1.2, 0) -- (2.2, 0);
            \draw [ultra thick, red, -<-=.60] (1.2, 0) -- (1.2, 0.75);
            \draw [ultra thick, red, -<-=.60] (1.2, .75) -- (1.2, 1.5);
 \filldraw[black] (1.2,0) circle (2pt);
 \node at (1.5, 1.4) {\footnotesize $W$};
 \node at (.5, 1.1) {\footnotesize $(-1)^F$};
 \draw [ ultra thick, blue, ->-=.6] (.2, 0.75) .. controls ( .7,.75) and (1.2,.75) .. (1.2, 1);
\draw [ ultra thick, blue, ->-=.6] (1.2, 0.5) .. controls (1.2, .75) and (1.8, .75) .. (2.3, 0.75);
\node at (1.2, -.3) {\footnotesize $\alpha$};
\end{tikzpicture}
\end{gathered}
&\,=\,\alpha_{4,2,3}^{-1}
\begin{gathered}
\begin{tikzpicture}[scale=1]
            \draw [ultra thick, densely dashed, ->-=.6] (.2, 0) -- (1.2, 0);
            \draw [ultra thick, densely dashed, ->-=.7] (1.2, 0) -- (2.2, 0);
            \draw [ultra thick, red, -<-=.60] (1.2, 0) -- (1.2, 0.75);
            \draw [ultra thick, red, -<-=.60] (1.2, .75) -- (1.2, 1.5);
 \filldraw[black] (1.2,0) circle (2pt);
 \node at (1.5, 1.4) {\footnotesize $W$};
 \node at (.5, 1.1) {\footnotesize $(-1)^F$};
 \draw [ ultra thick, blue, ->-=.6] (.2, 0.75) .. controls ( .7,.75) and (1.2,.75) .. (1.2, 1);
\draw [ ultra thick, blue, ->-=.6] (1.2, 0.5) .. controls (1.2, .75) and (1.8, .75) .. (2.3, 0.75);
\node at (1.15, -.3) {\footnotesize $\alpha-1$};
\filldraw[black] (1.2, 0.5) circle (1.5pt);
\end{tikzpicture}
\end{gathered}
\,=\,
-\frac{i}{2\alpha_{4,2,3}}\,
\begin{gathered}
\begin{tikzpicture}[scale=1]
            \draw [ultra thick, densely dashed, ->-=.6] (.2, 0) -- (1.2, 0);
            \draw [ultra thick, densely dashed, ->-=.7] (1.2, 0) -- (2.2, 0);
            \draw [ultra thick, red, -<-=.60] (1.2, 0) -- (1.2, 0.75);
            \draw [ultra thick, red, -<-=.60] (1.2, .75) -- (1.2, 1.5);
 \filldraw[black] (1.2,0) circle (2pt);
 \node at (1.5, 1.4) {\footnotesize $W$};
 \node at (.5, 1.1) {\footnotesize $(-1)^F$};
 \draw [ ultra thick, blue, ->-=.6] (.2, 0.75) .. controls ( .7,.75) and (1.2,.75) .. (1.2, 0.5);
\draw [ ultra thick, blue, ->-=.6] (1.2, 1) .. controls (1.2, .75) and (1.8, .75) .. (2.3, 0.75);
\node at (1.15, -.3) {\footnotesize $\alpha-1$};
 \filldraw[black] (1.2, 0.5) circle (1.5pt);
\end{tikzpicture}
\end{gathered}
\,-\,\frac{1}{2\alpha_{4,2,3}}\,
\begin{gathered}
\begin{tikzpicture}[scale=1]
            \draw [ultra thick, densely dashed, ->-=.6] (.2, 0) -- (1.2, 0);
            \draw [ultra thick, densely dashed, ->-=.7] (1.2, 0) -- (2.2, 0);
            \draw [ultra thick, red, -<-=.60] (1.2, 0) -- (1.2, 0.75);
            \draw [ultra thick, red, -<-=.60] (1.2, .75) -- (1.2, 1.5);
 \filldraw[black] (1.2,0) circle (2pt);
 \node at (1.5, 1.4) {\footnotesize $W$};
 \node at (.5, 1.1) {\footnotesize $(-1)^F$};
 \draw [ ultra thick, blue, ->-=.6] (.2, 0.75) .. controls ( .7,.75) and (1.2,.75) .. (1.2, 0.5);
\draw [ ultra thick, blue, ->-=.6] (1.2, 1) .. controls (1.2, .75) and (1.8, .75) .. (2.3, 0.75);
\node at (1.15, -.3) {\footnotesize $\alpha-1$};
 \filldraw[black] (1.2, 1) circle (1.5pt);
\end{tikzpicture}
\end{gathered}
\notag\\
&\,=\,
-\frac{i}{2}\frac{\beta_{2,4,3}}{\alpha_{4,2,3}}\,
\begin{gathered}
\begin{tikzpicture}[scale=1]
            \draw [ultra thick, densely dashed, ->-=.6] (.2, 0) -- (1.2, 0);
            \draw [ultra thick, densely dashed, ->-=.7] (1.2, 0) -- (2.2, 0);
            \draw [ultra thick, red, -<-=.60] (1.2, 0) -- (1.2, 0.75);
            \draw [ultra thick, red, -<-=.60] (1.2, .75) -- (1.2, 1.5);
 \filldraw[black] (1.2,0) circle (2pt);
 \node at (1.5, 1.4) {\footnotesize $W$};
 \node at (.5, 1.1) {\footnotesize $(-1)^F$};
 \draw [ ultra thick, blue, ->-=.6] (.2, 0.75) .. controls ( .7,.75) and (1.2,.75) .. (1.2, 0.5);
\draw [ ultra thick, blue, ->-=.6] (1.2, 1) .. controls (1.2, .75) and (1.8, .75) .. (2.3, 0.75);
\node at (1.2, -.3) {\footnotesize $\alpha$};
\end{tikzpicture}
\end{gathered}
\,-\,\frac{1}{2}\frac{\beta_{2,4,3}^2}{\alpha_{4,2,3}\alpha_{3,2,4}}\,
\begin{gathered}
\begin{tikzpicture}[scale=1]
            \draw [ultra thick, densely dashed, ->-=.6] (.2, 0) -- (1.2, 0);
            \draw [ultra thick, densely dashed, ->-=.7] (1.2, 0) -- (2.2, 0);
            \draw [ultra thick, red, -<-=.60] (1.2, 0) -- (1.2, 0.75);
            \draw [ultra thick, red, -<-=.60] (1.2, .75) -- (1.2, 1.5);
 \filldraw[black] (1.2,0) circle (2pt);
 \node at (1.5, 1.4) {\footnotesize $W$};
 \node at (.5, 1.1) {\footnotesize $(-1)^F$};
 \draw [ ultra thick, blue, ->-=.6] (.2, 0.75) .. controls ( .7,.75) and (1.2,.75) .. (1.2, 0.5);
\draw [ ultra thick, blue, ->-=.6] (1.2, 1) .. controls (1.2, .75) and (1.8, .75) .. (2.3, 0.75);
\node at (1.2, -.3) {\footnotesize $\alpha$};
 \filldraw[black] (1.2, 1) circle (1.5pt);
\end{tikzpicture}
\end{gathered}
\,=\,
\begin{gathered}
\begin{tikzpicture}[scale=1]
            \draw [ultra thick, densely dashed, ->-=.6] (.2, 0) -- (1.2, 0);
            \draw [ultra thick, densely dashed, ->-=.7] (1.2, 0) -- (2.2, 0);
            \draw [ultra thick, red, -<-=.60] (1.2, 0) -- (1.2, 0.75);
            \draw [ultra thick, red, -<-=.60] (1.2, .75) -- (1.2, 1.5);
 \filldraw[black] (1.2,0) circle (2pt);
 \node at (1.5, 1.4) {\footnotesize $W$};
 \node at (.5, 1.1) {\footnotesize $(-1)^F$};
 \draw [ ultra thick, blue, ->-=.6] (.2, 0.75) .. controls ( .7,.75) and (1.2,.75) .. (1.2, 0.5);
\draw [ ultra thick, blue, ->-=.6] (1.2, 1) .. controls (1.2, .75) and (1.8, .75) .. (2.3, 0.75);
\node at (1.2, -.3) {\footnotesize $\alpha$};
\end{tikzpicture}
\end{gathered}
\fe

\newpage
\appendix


\subsection*{Acknowledgements}
CC is supported by NSFC Grant No. 12575075. J.C. is supported by Fujian Provincial Natural Science Foundation of China (No.2025J01004), and the National Natural Science Foundation of China (Grants No.12247103).

\section{F-symbols associated to three $\mathbb{Z}_2$ classes}
\label{app: F-symbols}
In this section, we list the relevant F-symbols for the three different $\mathbb{Z}_2$ flavor symmetry, where the solid line represents the non-trivial $\mathbb{Z}_2$ topological defect line(TDL), and the dotted line a trivial TDL.

\subsection{$Z_2^{\nu_W=1,3,5,7}$}
\label{app:cq0}
There are \emph{four} inequivalent solutions which has been solved in \cite{Chang:2022hud}, and we list them here as the context self-contained.
\ie
\label{eq:cq0solution}
&\hspace{-.5cm}\begin{pmatrix}
\begin{gathered}
\begin{tikzpicture}[scale=.3]
\draw [ultra thick, red] (-1,1) -- (-2,2);
\draw [ultra thick, red] (-1,1) --(0,2);
\draw [ultra thick, red] (0,0) --(2,2);
\draw [ultra thick, red] (0,0)  -- (0,-1);
\draw [ultra thick, blue, densely dotted] (0,0) -- (-1,1) ;
\end{tikzpicture}
\end{gathered}
\\
\begin{gathered}
\begin{tikzpicture}[scale=.3]
\draw [ultra thick, red] (-1,1) -- (-2,2);
\draw [ultra thick, red] (-1,1) --(0,2);
\draw [ultra thick, red] (0,0) --(2,2);
\draw [ultra thick, red] (0,0)  -- (0,-1);
\draw [ultra thick, blue, densely dotted] (0,0) -- (-1,1) ;
\draw (0,0)\dotsolb {}{};
\draw (-1,1)\dotsolb {}{};
\end{tikzpicture}
\end{gathered}
\\
\begin{gathered}
\begin{tikzpicture}[scale=.3]
\draw [ultra thick, red] (-1,1) -- (-2,2);
\draw [ultra thick, red] (-1,1) --(0,2);
\draw [ultra thick, red] (0,0) --(2,2);
\draw [ultra thick, red] (0,0)  -- (0,-1);
\draw [ultra thick, blue, densely dotted] (0,0) -- (-1,1) ;
\draw (0,0)\dotsolb {}{};
\end{tikzpicture}
\end{gathered}
\\
\begin{gathered}
\begin{tikzpicture}[scale=.3]
\draw [ultra thick, red] (-1,1) -- (-2,2);
\draw [ultra thick, red] (-1,1) --(0,2);
\draw [ultra thick, red] (0,0) --(2,2);
\draw [ultra thick, red] (0,0)  -- (0,-1);
\draw [ultra thick, blue, densely dotted] (0,0) -- (-1,1) ;
\draw (-1,1)\dotsolb {}{};
\end{tikzpicture}
\end{gathered}
\end{pmatrix}
=
\begin{large}
\frac{\kappa}{\sqrt{2}}\begin{pmatrix}
1 & \gamma & 0 & 0 &\\
\gamma & -\gamma^2 & 0 & 0 & \\
0 & 0 & 1  & \gamma &  \\
0 & 0 & \gamma & -\gamma^2 &  \\
\end{pmatrix}
\end{large}
\cdot
\begin{pmatrix}
\begin{gathered}
\begin{tikzpicture}[scale=.3]
\draw [ultra thick, red] (0,0) -- (-2,2);
\draw [ultra thick, red] (1,1) --(0,2);
\draw [ultra thick, red] (1,1) --(2,2);
\draw [ultra thick, red] (0,0)  -- (0,-1);
\draw [ultra thick, blue, densely dotted] (0,0) -- (1,1) ;
\end{tikzpicture}
\end{gathered}
\\
\begin{gathered}
\begin{tikzpicture}[scale=.3]
\draw [ultra thick, red] (0,0) -- (-2,2);
\draw [ultra thick, red] (1,1) --(0,2);
\draw [ultra thick, red] (1,1) --(2,2);
\draw [ultra thick, red] (0,0)  -- (0,-1);
\draw [ultra thick, blue, densely dotted] (0,0) -- (1,1) ;
\draw (1,1)\dotsolb {}{};
\draw (0,0)\dotsolb {}{};
\end{tikzpicture}
\end{gathered}
\\
\begin{gathered}
\begin{tikzpicture}[scale=.3]
\draw [ultra thick, red] (0,0) -- (-2,2);
\draw [ultra thick, red] (1,1) --(0,2);
\draw [ultra thick, red] (1,1) --(2,2);
\draw [ultra thick, red] (0,0)  -- (0,-1);
\draw [ultra thick, blue, densely dotted] (0,0) -- (1,1) ;
\draw (0,0)\dotsolb {}{};
\end{tikzpicture}
\end{gathered}
\\
\begin{gathered}
\begin{tikzpicture}[scale=.3]
\draw [ultra thick, red] (0,0) -- (-2,2);
\draw [ultra thick, red] (1,1) --(0,2);
\draw [ultra thick, red] (1,1) --(2,2);
\draw [ultra thick, red] (0,0)  -- (0,-1);
\draw [ultra thick, blue, densely dotted] (0,0) -- (1,1) ;
\draw (1,1)\dotsolb {}{};
\end{tikzpicture}
\end{gathered}
\end{pmatrix}\,,
\fe
for $\kappa=\pm 1$ and $\gamma=e^{\frac{i \pi}{4}},\,e^{\frac{3i \pi}{4}}$.

\subsection{$Z_2^{\nu_W=2,6}$}
There are two inequivalent solutions has been solved in \cite{Chang:2022hud}. The corresponding F-symbols are 
\ie\label{eqn:hatC_F}
\begin{gathered}
\begin{tikzpicture}[scale=.75]
\draw [ultra thick, red] (-1,1) -- (-2,2);
\draw [ultra thick, red] (-1,1) --(0,2);
\draw [ultra thick, red] (0,0) --(2,2);
\draw [ultra thick, red] (0,0)  -- (0,-1);
\draw [ultra thick, blue, densely dotted] (0,0) -- (-1,1) ;
\draw (-1,1)\dotsolb {}{};
\end{tikzpicture}
\end{gathered}
\quad=\quad\kappa i
\begin{gathered}
\begin{tikzpicture}[scale=.75]
\draw [ultra thick, red] (-2,2) -- (0,0);
\draw [ultra thick, red] (0,2) --(1,1);
\draw [ultra thick, red] (1,1) --(2,2);
\draw [ultra thick, blue, densely dotted] (0,0)  -- (1,1);
\draw [ultra thick, red] (0,0) -- (0,-1) ;
\draw (1,1)\dotsolb {}{};
\end{tikzpicture}
\end{gathered}\,,
\fe
where $\kappa=\pm 1$. 

\subsection{$Z_2^{\nu_W=0,4}$}
There are two inequivalent solutions has been solved in \cite{Chang:2022hud}. The corresponding F-symbols are 
\ie\label{eqn:hatC_F}
\begin{gathered}
\begin{tikzpicture}[scale=.75]
\draw [ultra thick, red] (-1,1) -- (-2,2);
\draw [ultra thick, red] (-1,1) --(0,2);
\draw [ultra thick, red] (0,0) --(2,2);
\draw [ultra thick, red] (0,0)  -- (0,-1);
\draw [ultra thick, blue, densely dotted] (0,0) -- (-1,1) ;
\end{tikzpicture}
\end{gathered}
\quad=\quad\kappa 
\begin{gathered}
\begin{tikzpicture}[scale=.75]
\draw [ultra thick, red] (-2,2) -- (0,0);
\draw [ultra thick, red] (0,2) --(1,1);
\draw [ultra thick, red] (1,1) --(2,2);
\draw [ultra thick, blue, densely dotted] (0,0)  -- (1,1);
\draw [ultra thick, red] (0,0) -- (0,-1) ;
\end{tikzpicture}
\end{gathered}\,,
\fe
where $\kappa=\pm 1$. Note that the non-anomalous $\mathbb{Z}_2$ takes $\kappa=1$, which is the choice we take for our $Z$, i.e. $(-1)^F$ in this note.

\section{The fermion action rules in $Z_2^{\nu_W=1,3,5,7}\times Z_2^{\nu_W=0,4}$}
\label{appen:gauging}

Recall that the 1d Majorana fermion can freely move along the q-type objects. When it goes across a trivalent vertex consisted of one m-type and two q-type objects, we define the following fermion action rules, after gauge fixings:
\ie
\begin{gathered}
\begin{tikzpicture}[scale=.75]
\draw [ultra thick, red] (0,0)  -- (0,-1) ;
\draw [ultra thick,blue] (0,0) -- (.87,.5) ;
\draw [ultra thick,red] (0,0) -- (-.87,.5) ;
\draw (0,0)\dotsolb {right}{};
\node at (-1, 0.7) {\tiny{$q_1$}};
\node at (1, 0.7) {\tiny{$m$}};
\node at (0, -1.2) {\tiny{$q_2$}};
\end{tikzpicture}
\end{gathered}
\quad&=\quad\alpha_{q_1,m,q_2}^{-1}\begin{gathered}
\begin{tikzpicture}[scale=.75]
\draw [ultra thick,red] (0,0)  -- (0,-1) ;
\draw [ultra thick,blue] (0,0) -- (.87,.5) ;
\draw [ultra thick,red] (0,0) -- (-.87,.5) ;
\draw (-0.17,.1)\dotsolb {right}{};
\node at (-1, 0.7) {\tiny{$q_1$}};
\node at (1, 0.7) {\tiny{$m$}};
\node at (0, -1.2) {\tiny{$q_2$}};
\end{tikzpicture}
\end{gathered}\,,\qquad& \begin{gathered}
\begin{tikzpicture}[scale=.75]
\draw [ultra thick,red] (0,0)  -- (0,-1) ;
\draw [ultra thick,blue] (0,0) -- (.87,.5) ;
\draw [ultra thick,red] (0,0) -- (-.87,.5) ;
\draw (0,0)\dotsolb {right}{};
\node at (-1, 0.7) {\tiny{$q_1$}};
\node at (1, 0.7) {\tiny{$m$}};
\node at (0, -1.2) {\tiny{$q_2$}};
\end{tikzpicture}
\end{gathered}
\quad&=\quad
\alpha_{q_1,m,q_2}\begin{gathered}
\begin{tikzpicture}[scale=.75]
\draw [ultra thick,red] (0,0)  -- (0,-1) ;
\draw [ultra thick,blue] (0,0) -- (.87,.5) ;
\draw [ultra thick,red] (0,0) -- (-.87,.5) ;
\draw (0,-.2)\dotsolb {right}{};
\node at (-1, 0.7) {\tiny{$q_1$}};
\node at (1, 0.7) {\tiny{$m$}};
\node at (0, -1.2) {\tiny{$q_2$}};
\end{tikzpicture}
\end{gathered}\,,\notag\\
\begin{gathered}
\begin{tikzpicture}[scale=.75]
\draw [ultra thick,red] (0,0)  -- (0,-1) ;
\draw [ultra thick,red] (0,0) -- (.87,.5) ;
\draw [ultra thick,blue] (0,0) -- (-.87,.5) ;
\draw (0,0)\dotsolb {right}{};
\node at (-1, 0.7) {\tiny{$m$}};
\node at (1, 0.7) {\tiny{$q_1$}};
\node at (0, -1.2) {\tiny{$q_2$}};
\end{tikzpicture}
\end{gathered}
\quad&=\quad\beta_{m,q_1,q_2}^{-1}\begin{gathered}
\begin{tikzpicture}[scale=.75]
\draw [ultra thick,red] (0,0)  -- (0,-1) ;
\draw [ultra thick,red] (0,0) -- (.87,.5) ;
\draw [ultra thick,blue] (0,0) -- (-.87,.5) ;
\draw (0.17,.1)\dotsolb {right}{};
\node at (-1, 0.7) {\tiny{$m$}};
\node at (1, 0.7) {\tiny{$q_1$}};
\node at (0, -1.2) {\tiny{$q_2$}};
\end{tikzpicture}
\end{gathered}\,,\qquad& \begin{gathered}
\begin{tikzpicture}[scale=.75]
\draw [ultra thick,red] (0,0)  -- (0,-1) ;
\draw [ultra thick,red] (0,0) -- (.87,.5) ;
\draw [ultra thick,blue] (0,0) -- (-.87,.5) ;
\draw (0,0)\dotsolb {right}{};
\node at (-1, 0.7) {\tiny{$m$}};
\node at (1, 0.7) {\tiny{$q_1$}};
\node at (0, -1.2) {\tiny{$q_2$}};
\end{tikzpicture}
\end{gathered}
\quad&=\quad
\beta_{m,q_1,q_2}\begin{gathered}
\begin{tikzpicture}[scale=.75]
\draw [ultra thick,red] (0,0)  -- (0,-1) ;
\draw [ultra thick,red] (0,0) -- (.87,.5) ;
\draw [ultra thick,blue] (0,0) -- (-.87,.5) ;
\draw (0,-.2)\dotsolb {right}{};
\node at (-1, 0.7) {\tiny{$m$}};
\node at (1, 0.7) {\tiny{$q_1$}};
\node at (0, -1.2) {\tiny{$q_2$}};
\end{tikzpicture}
\end{gathered}\,,\notag\\
\begin{gathered}
\begin{tikzpicture}[scale=.75]
\draw [ultra thick,blue] (0,0)  -- (0,-1) ;
\draw [ultra thick,red] (0,0) -- (.87,.5) ;
\draw [ultra thick,red] (0,0) -- (-.87,.5) ;
\draw (0,0)\dotsolb {right}{};
\node at (-1, 0.7) {\tiny{$q_1$}};
\node at (1, 0.7) {\tiny{$q_2$}};
\node at (0, -1.2) {\tiny{$m$}};
\end{tikzpicture}
\end{gathered}
\quad&=\quad\gamma_{q_1,q_2,m}\begin{gathered}
\begin{tikzpicture}[scale=.75]
\draw [ultra thick,blue] (0,0)  -- (0,-1) ;
\draw [ultra thick,red] (0,0) -- (.87,.5) ;
\draw [ultra thick,red] (0,0) -- (-.87,.5) ;
\draw (-0.17,.1)\dotsolb {right}{};
\node at (-1, 0.7) {\tiny{$q_1$}};
\node at (1, 0.7) {\tiny{$q_2$}};
\node at (0, -1.2) {\tiny{$m$}};
\end{tikzpicture}
\end{gathered}\,,\qquad& \begin{gathered}
\begin{tikzpicture}[scale=.75]
\draw [ultra thick,blue] (0,0)  -- (0,-1) ;
\draw [ultra thick,red] (0,0) -- (.87,.5) ;
\draw [ultra thick,red] (0,0) -- (-.87,.5) ;
\draw (0,0)\dotsolb {right}{};
\node at (-1, 0.7) {\tiny{$q_1$}};
\node at (1, 0.7) {\tiny{$q_2$}};
\node at (0, -1.2) {\tiny{$m$}};
\end{tikzpicture}
\end{gathered}
\quad&=\quad
\gamma_{q_1,q_2,m}^{-1}\begin{gathered}
\begin{tikzpicture}[scale=.75]
\draw [ultra thick,blue] (0,0)  -- (0,-1) ;
\draw [ultra thick,red] (0,0) -- (.87,.5) ;
\draw [ultra thick,red] (0,0) -- (-.87,.5) ;
\draw (0.17,.1)\dotsolb {right}{};
\node at (-1, 0.7) {\tiny{$q_1$}};
\node at (1, 0.7) {\tiny{$q_2$}};
\node at (0, -1.2) {\tiny{$m$}};
\end{tikzpicture}
\end{gathered}\,.
\label{eq:fermion_action}
\fe
Respect to the fusion rules, we have the following 12 factors, grouped into three sets 
\ie
&U_1=\{\alpha_{3,1,3},\, \beta_{1,3,3},\, \gamma_{3,3,1}\}\notag\\
&U_2=\{\alpha_{4,1,4},\, \beta_{1,4,4},\, \gamma_{4,4,1}\}\notag\\
&U_3=\{\alpha_{3,2,4},\, \alpha_{4,2,3},\, \beta_{2,3,4},\,\beta_{2,4,3},\, \gamma_{3,4,2},\,\gamma_{4,3,2}\}\,,
\fe
where we used numeric labels to denote the TDLs as in \eqref{eqn:line_numeric_label}. Since both $\{I,\, W\}$ and $\{I,\, WZ\}$ form super-fusion categories isomorphic to the fermionic Ising categories, the coefficients in the sets $U_1$ and $U_2$ are given by the universal sectors of the two superfusion subcatories \cite{Chang:2022hud},
\ie
\alpha_{3,1,3}=\alpha_{4,1,4}=\beta_{1,3,3}=\beta_{1,4,4}=1\,,\ \ \gamma_{3,3,1},\, \gamma_{4,4,1}\in \left\{e^{\frac{\pi i}{4}},\, e^{\frac{3\pi i}{4}}\right\}\,.
\fe
For each of the four combinations of $\gamma_{3,3,1}$ and $ \gamma_{4,4,1}$, one can further find 4 gauge inequivalent solutions to the super-pentagon equations. Therefore, we have overall 16 solutions. For later convenience, we here list all four solution sets to $U_3$ for the four gauge inequivalent choices of $\{\gamma_{3,3,1},\,\gamma_{4,4,1}\}$
\begin{table}[h!]
\centering
\footnotesize
\begin{tabular}{|c||c|c|c|c|c|c|}
\hline
 $(\gamma_{3,3,1},\,\gamma_{4,4,1})$ & $\alpha_{3,2,4}$ & $\alpha_{4,2,3}$ & $\beta_{2,3,4}$  & $\beta_{2,4,3}$ & $\gamma_{3,4,2}$ & $\gamma_{4,3,2}$
\\\hline\hline
$(e^{\frac{\pi i}{4}},\,e^{\frac{\pi i}{4}})$ & $1$ & $1$ & $1$ & $1$ & $e^{\frac{\pi i}{4}}$ & $e^{\frac{\pi i}{4}}$ \\\hline
$(e^{\frac{\pi i}{4}},\,e^{\frac{3\pi i}{4}})$ & $1$ & $1$ & $i$ & $i$ & $e^{\frac{\pi i}{4}}$ & $e^{\frac{3\pi i}{4}}$ \\\hline
$(e^{\frac{3\pi i}{4}},\,e^{\frac{\pi i}{4}})$ & $1$ & $1$ & $i$ & $i$ & $e^{\frac{3\pi i}{4}}$ & $e^{\frac{\pi i}{4}}$ \\\hline
$(e^{\frac{3\pi i}{4}},\,e^{\frac{3\pi i}{4}})$ & $1$ & $1$ & $1$ & $1$ & $e^{\frac{3\pi i}{4}}$ & $e^{\frac{3\pi i}{4}}$ 
\\\hline
\end{tabular}
\end{table}

Now, let us impose the condition \eqref{eqn:1dMF_passing_Fparity}. The quartic vertex can be resolved into a pair of cubic vertices as
\ie
\begin{gathered}
\begin{tikzpicture}[scale=1]
\draw [line,lightgray] (0,0)  -- (0,2) -- (2,2) -- (2,0) -- (0,0) ;
\draw [ultra thick,red] (1,0) -- (1,2) ;
\draw [ultra thick, blue] (1,.7) .. controls (1,1) and (1.5,1) .. (2,1)
(0,1) .. controls (.5,1) and (1,1) .. (1,1.3) ;
\draw (1,0.5)\dotsol {right}{}; 
\node at (0.3,1.15) {\tiny $2$};
\node at (1.7,0.85) {\tiny $2$};
\node at (1.3,1.1) {\tiny $4(3)$};
\node at (1.3,1.8) {\tiny $3(4)$};
\node at (1.3,0.2) {\tiny $3(4)$};

\end{tikzpicture}
\end{gathered}
\quad\quad {\rm or} \quad\quad
\begin{gathered}
\begin{tikzpicture}[scale=1]
\draw [line,lightgray] (0,0)  -- (0,2) -- (2,2) -- (2,0) -- (0,0) ;
\draw [ultra thick, red] (1,0) -- (1,2) ;
\draw [ultra thick, blue] (1,1.3) .. controls (1,1) and (1.5,1) .. (2,1)
(0,1) .. controls (.5,1) and (1,1) .. (1,.7) ;
\draw (1,0.5)\dotsol {right}{}; 
\node at (0.3,1.15) {\tiny $2$};
\node at (1.7,0.85) {\tiny $2$};
\node at (1.3,0.90) {\tiny $4(3)$};
\node at (1.3,1.8) {\tiny $3(4)$};
\node at (1.3,0.2) {\tiny $3(4)$};
\end{tikzpicture}
\end{gathered}
\,,
\label{graph:FZM_-1F}
\fe
which are related by F-moves. Now, moving the Majorana mode along the q-type lines from bottom to top, the two graphs will pick up a series factors as $\alpha_{4,2,3}^{-2}\beta_{2,3,4}^{-2}$ or $\alpha_{3,2,4}^{-2}\beta_{2,4,3}^{-2}$. We thus requires that
\ie
\alpha_{4,2,3}^{-2}\beta_{2,3,4}^{-2}=\alpha_{3,2,4}^{-2}\beta_{2,4,3}^{-2}=-1\,.
\label{eq:criteria}
\fe
Scaning the 16 solutions, one finds that there are 8 solutions with
\ie
\left(\gamma_{3,3,1},\, \gamma_{4,4,1}\right)=\left(e^{\frac{\pi i}{4}},\, e^{\frac{3\pi i}{4}}\right)\,, \ \ {\rm or}\ \ \left(e^{\frac{3\pi i}{4}},\, e^{\frac{\pi i}{4}}\right)\,,
\fe
satisfying the criteria, while the others give ``+1" and are excluded.

\section{ A glimpse on 2D Majorana Fermions}
\label{appen:2dMF}
In this section, we recall some basic properties of $v$ Copies of 2d Majorana Fermions. Let us start with a single one, whose action reads
\begin{equation}
    S_{maj}=\int d^2x \bar{\chi}\slashed{D}\chi
\end{equation}
This theory admits a $(-1)^F$ symmetry and $(-1)^{F_L}$ which acts as 
\[
\begin{split}
    (-1)^F:\quad & \chi\rightarrow -\chi \\
    (-1)^{F_L}:\quad &\chi \rightarrow \gamma_3\chi \qquad (\textbf{Lorentz sigature})
    \end{split}
\]
the torus partition function in the NS-NS sector reads
\begin{align}
    \mathcal Z_{1}=|\chi_0+\chi_{1/2}|^{2}\,.
\end{align}
where $\chi_h$ denotes the character of a Virasoro module of conformal weight $h$, which is given by
\begin{equation}
    \chi_0(\tau)=\frac{1}{2}(\sqrt{\frac{\theta_3}{\eta}}+\sqrt{\frac{\theta_4}{\eta}}), \quad  \chi_{\frac{1}{2}}(\tau)=\frac{1}{2}(\sqrt{\frac{\theta_3}{\eta}}-\sqrt{\frac{\theta_4}{\eta}}), \quad  \chi_{\frac{1}{16}}=\frac{1}{\sqrt{2}}(\sqrt{\frac{\theta_2}{\eta}}) 
\end{equation}
Generalizing to $\nu$ copies, the partition function is then 
\begin{align}
    \mathcal Z_{\nu}=|\chi_0+\chi_{1/2}|^{2\nu}\,.
\end{align}
In each copy of 2d Majorana fermion, there is a $(-1)^{F_L}$ symmetry, together with the univeral $(-1)^F$ symmetry, which acts on these chracters as 
\ie
(-1)^{F_L}:\quad \chi_0\rightarrow \chi_0\,,\quad \chi_{\frac{1}{2}}\rightarrow -\chi_{\frac{1}{2}},\,\notag\\
\fe

\subsection{Majorana Fermion as Fermionization of Ising CFT: Duality $\mathcal N$-line from fermionic Ising}
It has been known that the Majorana fermion can be viewed as ferminzation of Ising CFT \cite{Karch:2019lnn}. In this subsection, we establish a relation between the Kramers-Wannier duality line in the bosonic Ising and various TDLs in the fermionized one on the level of the partition functions.

First, recall that the fermionic theory can be obtained by stacking a Kitaev chain onto the Ising model and then gauging the diagonal $\mathbb Z_2$ of $\mathbb Z_{2,b}\times \mathbb Z_{2,f}$, where the subscripts ``b" and ``f" label the $\mathbb Z_2$ symmetry of Ising and Kitaev chain respectively. In terms of partition functions, one finds that
\begin{align}
Z_{\rm Maj.}[\rho]=\frac{1}{\sqrt{\left|H^1(M, G)\right|}}\sum_{a\in H^1(M,\mathbb Z_2)}(-1)^{{\rm Arf}(\rho+a)+{\rm Arf}(\rho)}Z_{\rm Ising}[a]\,,
\label{eq:fermionization_Arf2}
\end{align}
or on $\mathbb T^2$,
\begin{align}
Z_{\rm Maj.}[\rho_1,\rho_2]=\frac{1}{2}\sum_{a_1,a_2\in \mathbb Z_2}(-1)^{(\rho_1+a_1)(\rho_2+a_2)+\rho_1\rho_2}Z_{\rm Ising}[a_1,a_2]\,,
\end{align}
where $\rho$ and $a$ denote the spin-structure and the $\mathbb Z_2$ gauge field. Using above equations, one can rewrite the bosonic partion function in terms of the fermionic one, e.g.
\begin{align}
Z_{\rm Ising}[0,0]=\frac{1}{2}\left(Z_{\rm Maj.}[0,0]+Z_{\rm Maj.}[0,1]+Z_{\rm Maj.}[1,0]+Z_{\rm Maj.}[1,1]\right)\,,
\end{align}
where $Z_{\rm Ising}[0,0]$ stands for the Ising partition function without insertion of any TDL, and $Z_{\rm Maj.}[i,j]$ for fermionic partition function under NS or Ramond boundary conditions for $i,j=0$ or $1$ along temporal or spacial directions. Instead, one can always fix the boundary conditions to be NS type, and insert the $(-1)^F$-line to change them to Ramond type. In this picture, we rewrite above equation as
\begin{align}
\begin{gathered}
\begin{tikzpicture}[scale=1]
\draw [line,lightgray] (0,0)  -- (0,2) -- (2,2) -- (2,0) -- (0,0) ;
\node at (1.8, 0.2) {\footnotesize b};
\end{tikzpicture}
\end{gathered}
=\frac{1}{2}\left(\ 
\begin{gathered}
\begin{tikzpicture}[scale=1]
\draw [line,lightgray] (0,0)  -- (0,2) -- (2,2) -- (2,0) -- (0,0) ;
\node at (1.8, 0.2) {\footnotesize f};
\end{tikzpicture}
\end{gathered}
+
\begin{gathered}
\begin{tikzpicture}[scale=1]
\draw [line,lightgray] (0,0)  -- (0,2) -- (2,2) -- (2,0) -- (0,0) ;
\draw [ultra thick, blue] (1,0) -- (1,2) ;
\node at (1.8, 0.2) {\footnotesize f};
\end{tikzpicture}
\end{gathered}
+
\begin{gathered}
\begin{tikzpicture}[scale=1]
\draw [line,lightgray] (0,0)  -- (0,2) -- (2,2) -- (2,0) -- (0,0) ;
\draw [ultra thick,blue] (0,1) -- (2,1) ;
\node at (1.8, 0.2) {\footnotesize f};
\end{tikzpicture}
\end{gathered}
+
\begin{gathered}
\begin{tikzpicture}[scale=1]
\draw [line,lightgray] (0,0)  -- (0,2) -- (2,2) -- (2,0) -- (0,0) ;
\draw [ultra thick,blue] (0,1) -- (2,1) ;
\draw [ultra thick,blue] (1,0) -- (1,2) ;
\node at (1.8, 0.2) {\footnotesize f};
\end{tikzpicture}
\end{gathered}
\ \right)\,,
\label{eq:Maj_to_Ising}
\end{align}
where the blue line denotes the $(-1)^F$-line, subscript ``b" stands for the bosonic theory, and ``f" for the fermionic one under the NS boundary condition along both space and time directions. Eq.\,(\ref{eq:Maj_to_Ising}) explicitly tells us that the Ising model can be obtained by gauging the $(-1)^F$-line of the Majorana one.

Now let us insert the duality $\mathcal N$-line in the bosonic theory along temporal direction. It will define a defect Hilbert space $\mathcal H_{\mathcal N}$, on which the partition function is evaluated. One may ask how the eq.\,(\ref{eq:Maj_to_Ising}) get modified? Since we have known that, after gauging the $\mathbb Z_2$ $W$-line on the bosnic side, the $\mathcal N$-line turns out to be a q-type TDL in the Majorana theory. It hints us that we should correspondingly insert a $(-1)^{F_L}$ or  $(-1)^{F_R}$ line on the RHS of eq.\,(\ref{eq:Maj_to_Ising}). Therefore we propose the following equation:
\begin{align}
\begin{gathered}
\begin{tikzpicture}[scale=1]
\draw [line,lightgray] (0,0)  -- (0,2) -- (2,2) -- (2,0) -- (0,0) ;
\draw [ultra thick] (1,0) -- (1,2) ;
\node at (1.3,1.1) {\tiny $\mathcal N$};
\node at (1.8, 0.2) {\footnotesize b};
\end{tikzpicture}
\end{gathered}
=\frac{1}{\sqrt{2}}\left(\ 
\begin{gathered}
\begin{tikzpicture}[scale=1]
\draw [line,lightgray] (0,0)  -- (0,2) -- (2,2) -- (2,0) -- (0,0) ;
\draw [ultra thick,red] (1,0) -- (1,2) ;
\node at (1.5,1.2) {\tiny $(-1)^{F_L}$};
\node at (1.8, 0.2) {\footnotesize f};
\end{tikzpicture}
\end{gathered}
+
\begin{gathered}
\begin{tikzpicture}[scale=1]
\draw [line,lightgray] (0,0)  -- (0,2) -- (2,2) -- (2,0) -- (0,0) ;
\draw [ultra thick,red] (1,0) -- (1,2) ;
\node at (1.5,1.2) {\tiny $(-1)^{F_R}$};
\node at (1.8, 0.2) {\footnotesize f};
\end{tikzpicture}
\end{gathered}
+
\begin{gathered}
\begin{tikzpicture}[scale=1]
\draw [line,lightgray] (0,0)  -- (0,2) -- (2,2) -- (2,0) -- (0,0) ;
\draw [ultra thick,blue] (0,1) -- (2,1) ;
\draw [ultra thick,red] (1,0) -- (1,2) ;
\node at (1.5,1.4) {\tiny $(-1)^{F_L}$};
\node at (0.47,0.8) {\tiny $(-1)^{F}$};
\node at (1.8, 0.2) {\footnotesize f};
\end{tikzpicture}
\end{gathered}
+
\begin{gathered}
\begin{tikzpicture}[scale=1]
\draw [line,lightgray] (0,0)  -- (0,2) -- (2,2) -- (2,0) -- (0,0) ;
\draw [ultra thick,blue] (0,1) -- (2,1) ;
\draw [ultra thick,red] (1,0) -- (1,2) ;
\node at (1.5,1.4) {\tiny $(-1)^{F_R}$};
\node at (0.47,0.8) {\tiny $(-1)^{F}$};
\node at (1.8, 0.2) {\footnotesize f};
\end{tikzpicture}
\end{gathered}
\ \right)\,,
\label{eq:Maj_to_Ising_with_N}
\end{align}
where the prefactor $\frac{1}{\sqrt{2}}$ is due to the unusual normalization of the q-type TDL $(-1)^{F_{L,R}}$ \cite{Chang:2022hud}.


In the rest of the section, we will show that the eq.\,(\ref{eq:Maj_to_Ising_with_N}) is indeed held. To this end, we first show that the partition functions corresponding to the last two terms in the RHS of eq.\,(\ref{eq:Maj_to_Ising_with_N}) vanish. Notice that the last two diagrams, after resolving 4-way junctions, are precisely the diagrams we discussed in (\ref{graph:FZM_-1F}). Now let us focus on the third diagram and consider pair-created one-dimensional Majorana fermion on the $(-1)^{F_L}$-line as follows:
\begin{align}
\begin{gathered}
\begin{tikzpicture}[scale=1]
\draw [line,lightgray] (0,0)  -- (0,2) -- (2,2) -- (2,0) -- (0,0) ;
\draw [ultra thick,blue] (0,1) -- (2,1) ;
\draw [ultra thick,red] (1,0) -- (1,2) ;
\node at (1.5,1.4) {\tiny $(-1)^{F_L}$};
\node at (0.47,0.8) {\tiny $(-1)^{F}$};
\draw (1,0.8)\dotsolb {right}{\footnotesize 1}; 
\draw (1,0.4)\dotsolb {right}{\footnotesize 2}; 
\node at (1.8, 0.2) {\footnotesize f};
\end{tikzpicture}
\end{gathered}
\quad=\quad
-\quad
\begin{gathered}
\begin{tikzpicture}[scale=1]
\draw [line,lightgray] (0,0)  -- (0,2) -- (2,2) -- (2,0) -- (0,0) ;
\draw [ultra thick,blue] (0,1) -- (2,1) ;
\draw [ultra thick,red] (1,0) -- (1,2) ;
\node at (1.5,1.4) {\tiny $(-1)^{F_L}$};
\node at (0.47,0.8) {\tiny $(-1)^{F}$};
\draw (1,1.8)\dotsolb {right}{\footnotesize 1}; 
\draw (1,0.4)\dotsolb {right}{\footnotesize 2}; 
\node at (1.8, 0.2) {\footnotesize f};
\end{tikzpicture}
\end{gathered}
\quad=\quad
+\quad
\begin{gathered}
\begin{tikzpicture}[scale=1]
\draw [line,lightgray] (0,0)  -- (0,2) -- (2,2) -- (2,0) -- (0,0) ;
\draw [ultra thick,blue] (0,1) -- (2,1) ;
\draw [ultra thick,red] (1,0) -- (1,2) ;
\node at (1.5,1.4) {\tiny $(-1)^{F_L}$};
\node at (0.47,0.8) {\tiny $(-1)^{F}$};
\draw (1,0.05)\dotsolb {right}{\footnotesize 1}; 
\draw (1,0.4)\dotsolb {right}{\footnotesize 2}; 
\node at (1.8, 0.2) {\footnotesize f};
\end{tikzpicture}
\end{gathered}
\quad=\quad
-\quad
\begin{gathered}
\begin{tikzpicture}[scale=1]
\draw [line,lightgray] (0,0)  -- (0,2) -- (2,2) -- (2,0) -- (0,0) ;
\draw [ultra thick,blue] (0,1) -- (2,1) ;
\draw [ultra thick,red] (1,0) -- (1,2) ;
\node at (1.5,1.4) {\tiny $(-1)^{F_L}$};
\node at (0.47,0.8) {\tiny $(-1)^{F}$};
\draw (1,0.8)\dotsolb {right}{\footnotesize 1}; 
\draw (1,0.4)\dotsolb {right}{\footnotesize 2}; 
\node at (1.8, 0.2) {\footnotesize f};
\end{tikzpicture}
\end{gathered},
\end{align}
where $-1$ in the first equality stems from the fact the first 1d fermions get across $(-1)^F$, the $-1$ in the third equality comes from the exchange of the two fermions, and in the second equality we employ the fact that rotating a 1d fermion by $\pi$ results in a minus sign in the NS boundary condition. Similiarly, we can run the same argument for the fourth diagram    in the RHS of eq.\,(\ref{eq:Maj_to_Ising_with_N}), hence both of them contribute trivial to the partition function. This result generalizes the fact that, for the Majorana fermion, the RR sector partition function is zero due to the appearance of zero mode.

Now we are left to show
\begin{align}
\begin{gathered}
\begin{tikzpicture}[scale=1]
\draw [line,lightgray] (0,0)  -- (0,2) -- (2,2) -- (2,0) -- (0,0) ;
\draw [ultra thick] (1,0) -- (1,2) ;
\node at (1.3,1.1) {\tiny $\mathcal N$};
\node at (1.8, 0.2) {\footnotesize b};
\end{tikzpicture}
\end{gathered}
\quad=\quad
\frac{1}{\sqrt{2}}\quad
\begin{gathered}
\begin{tikzpicture}[scale=1]
\draw [line,lightgray] (0,0)  -- (0,2) -- (2,2) -- (2,0) -- (0,0) ;
\draw [ultra thick,red] (1,0) -- (1,2) ;
\node at (1.5,1.2) {\tiny $(-1)^{F_L}$};
\node at (1.8, 0.2) {\footnotesize f};
\end{tikzpicture}
\end{gathered}
\quad+\quad
\frac{1}{\sqrt{2}}\quad
\begin{gathered}
\begin{tikzpicture}[scale=1]
\draw [line,lightgray] (0,0)  -- (0,2) -- (2,2) -- (2,0) -- (0,0) ;
\draw [ultra thick,red] (1,0) -- (1,2) ;
\node at (1.5,1.2) {\tiny $(-1)^{F_R}$};
\node at (1.8, 0.2) {\footnotesize f};
\end{tikzpicture}
\end{gathered}
\,.
\end{align}
First we write down the partition functoin of the LHS of above equation,
\begin{align}
Z_{b,\,\mathcal N}=\Tr_{\mathcal H_{\mathcal N}}\left(q^{L_0-\frac{c}{24}}\bar q^{\bar L_0-\frac{c}{24}}\right)=\left(\chi_0+\chi_{1/2}\right)\bar\chi_{1/16}+h.c.\,.
\end{align}
 Next we will compute the partition function for the graph
\begin{align}
\begin{gathered}
\begin{tikzpicture}[scale=1]
\draw [line,lightgray] (0,0)  -- (0,2) -- (2,2) -- (2,0) -- (0,0) ;
\draw [ultra thick,red] (0,1) -- (2,1) ;
\node at (1.5,1.2) {\tiny $(-1)^{F_L}$};
\node at (1.8, 0.2) {\footnotesize f};
\end{tikzpicture}
\end{gathered}
\label{graph:-1FL_space}
\end{align}
Recall that, without insertion of $(-1)^{F_L}$, the fermionic partition function is just for NS-NS sector, and spells as
\begin{align}
Z_{f\, {\rm NS}}\equiv\Tr_{\rm NSNS}\left(q^{L_0-\frac{c}{24}}\bar q^{\bar L_0-\frac{c}{24}}\right)=\left(\chi_0+\chi_{1/2}\right)\left(\bar\chi_0+\bar\chi_{1/2}\right)
\end{align}
With insertion of the $(-1)^{F_L}$ along the spacial direction, we need to consider how the $(-1)^{F_L}$ acts on various states. Notice that the $(-1)^{F_L}$-line is only sensitive to the anti-holomorphic piece and changes sign for the character of $\bar\chi_{1/2}$, while keeping the holomorphic one intact. We thus have
\begin{align}
Z_{f\, {\rm NS}}[(-1)^{F_L}]\equiv\Tr_{\rm NSNS}\left((-1)^{F_L}q^{L_0-\frac{c}{24}}\bar q^{\bar L_0-\frac{c}{24}}\right)=\left(\chi_0+\chi_{1/2}\right)\left(\bar\chi_0-\bar\chi_{1/2}\right)\,,
\end{align}
i.e. the partition function for graph (\ref{graph:-1FL_space}). Now we apply a S-transformation to obtain the graph with insertion of the $(-1)^{F_L}$ along temporal direction,
\begin{align}
\begin{gathered}
\begin{tikzpicture}[scale=1]
\draw [line,lightgray] (0,0)  -- (0,2) -- (2,2) -- (2,0) -- (0,0) ;
\draw [ultra thick,red] (0,1) -- (2,1) ;
\node at (1.5,1.2) {\tiny $(-1)^{F_L}$};
\node at (1.8, 0.2) {\footnotesize f};
\end{tikzpicture}
\end{gathered}
\quad\longrightarrow\quad
\begin{gathered}
\begin{tikzpicture}[scale=1]
\draw [line,lightgray] (0,0)  -- (0,2) -- (2,2) -- (2,0) -- (0,0) ;
\draw [ultra thick,red] (1,0) -- (1,2) ;
\node at (1.5,1.2) {\tiny $(-1)^{F_L}$};
\node at (1.8, 0.2) {\footnotesize f};
\end{tikzpicture}
\end{gathered}
\,.
\end{align}
Correspondingly the characters transform as
\begin{align}
\chi_0+\chi_{1/2}\ \longrightarrow\ \chi_0+\chi_{1/2}\,\ \ {\rm and}\ \ \bar\chi_0-\bar\chi_{1/2}\ \longrightarrow\ \sqrt{2}\bar\chi_{1/16}\,.
\end{align}
Therefore we precisely obtain
\begin{align}
\frac{1}{\sqrt{2}}\quad
\begin{gathered}
\begin{tikzpicture}[scale=1]
\draw [line,lightgray] (0,0)  -- (0,2) -- (2,2) -- (2,0) -- (0,0) ;
\draw [ultra thick,red] (1,0) -- (1,2) ;
\node at (1.5,1.2) {\tiny $(-1)^{F_L}$};
\node at (1.8, 0.2) {\footnotesize f};
\end{tikzpicture}
\end{gathered}
\quad=\quad
\left(\chi_0+\chi_{1/2}\right)\bar\chi_{1/16}\,.
\end{align}
Similarly one can show that
\begin{align}
\frac{1}{\sqrt{2}}\quad
\begin{gathered}
\begin{tikzpicture}[scale=1]
\draw [line,lightgray] (0,0)  -- (0,2) -- (2,2) -- (2,0) -- (0,0) ;
\draw [ultra thick,red] (1,0) -- (1,2) ;
\node at (1.5,1.2) {\tiny $(-1)^{F_R}$};
\node at (1.8, 0.2) {\footnotesize f};
\end{tikzpicture}
\end{gathered}
\quad=\quad
\left(\bar\chi_0+\bar\chi_{1/2}\right)\chi_{1/16}\,.
\end{align}
Overall we proved the eq.\,(\ref{eq:Maj_to_Ising_with_N}).

In addition, one can also find that the bosonic partition function with a $\eta$ $\mathbb{Z}_2$-line inserted on the temporal direction reads 
\begin{align}
\begin{gathered}
\begin{tikzpicture}[scale=1]
\draw [line,lightgray] (0,0)  -- (0,2) -- (2,2) -- (2,0) -- (0,0) ;
\draw [ultra thick] (1,0) -- (1,2) ;
\node at (1.3,1.1) {\footnotesize{$\eta$}};
\node at (1.8, 0.2) {\footnotesize b};
\end{tikzpicture}
\end{gathered}
=\frac{1}{2}\left(\ 
\begin{gathered}
\begin{tikzpicture}[scale=1]
\draw [line,lightgray] (0,0)  -- (0,2) -- (2,2) -- (2,0) -- (0,0) ;
\node at (1.8, 0.2) {\footnotesize f};
\end{tikzpicture}
\end{gathered}
+
\begin{gathered}
\begin{tikzpicture}[scale=1]
\draw [line,lightgray] (0,0)  -- (0,2) -- (2,2) -- (2,0) -- (0,0) ;
\draw [ultra thick,blue] (1,0) -- (1,2) ;
\node at (1.8, 0.2) {\footnotesize f};
\end{tikzpicture}
\end{gathered}
-
\begin{gathered}
\begin{tikzpicture}[scale=1]
\draw [line,lightgray] (0,0)  -- (0,2) -- (2,2) -- (2,0) -- (0,0) ;
\draw [ultra thick,blue] (0,1) -- (2,1) ;
\node at (1.8, 0.2) {\footnotesize f};
\end{tikzpicture}
\end{gathered}
-
\begin{gathered}
\begin{tikzpicture}[scale=1]
\draw [line,lightgray] (0,0)  -- (0,2) -- (2,2) -- (2,0) -- (0,0) ;
\draw [ultra thick,blue] (0,1) -- (2,1) ;
\draw [ultra thick,blue] (1,0) -- (1,2) ;
\node at (1.8, 0.2) {\footnotesize f};
\end{tikzpicture}
\end{gathered}
\ \right)\,,
\label{eq:Maj_to_Ising_eta}
\end{align}
It is easy to verify that eq.\,(\ref{eq:Maj_to_Ising_with_N}) can be also obtained by fusing $\mathcal N$-line on the LHS and $(-1)^{F_L}$-line on the RHS of eq.\,(\ref{eq:Maj_to_Ising_eta}), by noticing that $\mathcal N\cdot \eta=\mathcal N$.

\section{Fermionic number operator $F$}
\label{appix:fermionnumber}
If a Lagrangian with Dirac fermions is invariant under a global $U(1)$ symmetry 
\[
\psi \rightarrow e^{i\alpha}\psi 
\]
then one can define the (normal order) \footnote{Otherwise the naive version of Fermion number $N=\int d^d x \psi^\dag\psi$ is divergent due to the Dirac sea.} fermion number associated with the conserved current 
as
\[
F=\frac{1}{2}\int d^dx [\psi^\dag, \psi]
\]
By choosing a complete orthonomal basis of the solution of Dirac Hamiltonian $H \psi=E \psi$, we can decompose 
\[
\psi=\sum_{n, E_n>0} a_n u_n+\sum_{m, E_m<0} c_m v_m
\]
where we further split the basis into the postive- and negative-energy basis $(u_n, v_n)$.  And $a_n$ and $c_m$ denotes the annihilator operators of positive- and negative fermions. In terms of the creation and annihilation operators, this can be expressed as 
\[
F=\sum_{E_n>0} (a_n^\dag a_n-\frac{1}{2})+\sum_{E_n<0} (c_m^\dag c_m-\frac{1}{2})+N_{0}
\]
where $N_0:=a_0^\dag a_0-\frac{1}{2}$ denotes the zero modes contribution. Now we note that annihilating a negative-energy fermion is equivalent to creating an postive-energy antiparticle(hole), which renders us to define the antiparticle(hole) operators $b_m, b_m^\dag$ as 
\[
b_m^\dag=c_m, \qquad b_m=c_m^\dag
\]
Therefore, we conclude that 
\begin{equation}
    \begin{split}
F=&\sum_{E_n>0}(a_n^\dag a_n-\frac{1}{2})+\sum_{E_n<0} (c_m^\dag c_m-\frac{1}{2})+N_{0} \\
=&\sum_{E_n>0}(a_n^\dag a_n-\frac{1}{2})+\sum_{E_n<0} (b_m b_m^\dag-\frac{1}{2})+N_{0}\\
=&\sum_{E_n>0}a_n^\dag a_n-\sum_{E_n<0}b_m^\dag b_m-\frac{1}{2}(\sum_{E_n>0}1-\sum_{E_m<0}1)+N_{0}\\
=&\sum_{E_n>0}a_n^\dag a_n-\sum_{E_n<0}b_m^\dag b_m-\frac{1}{2}\eta(0)+N_{0}\\
    \end{split}
\end{equation}
where the last term $\frac{\eta(0)}{2}$ is essentially the spectral asymmetry of the system. Note that the minus sign in the second term above reflects the fact that creating an antiparticle (i.e. a hole in Dirace sea) reduces the fermion number by one unit.

When evaluated in a vacuum, the above reduces to the last two terms as $a_p, b_p$ annilate the vaucum state
\[
\langle N\rangle_v=-\frac{1}{2}\eta(0)+\langle N_0\rangle_v
\]
 Furthermore, for a free fermions with no interactions, this is exactly zero. Now let us consider backgrounds with kinks/solitons in 2d. 

\subsection*{1. The \texorpdfstring{$N=(1,1)$}{N=(1,1)} Case: One Real Majorana Zero Mode}

There is no global U(1) symmetry associated to Majorana fermions $\psi$ in 2d $N=(1,1)$ theories. Nevertheless, there is a discrete unbroken $\mathbb{Z}_2$ symmetry which can be thought of acting as 
\[
\psi \rightarrow -\psi
\]
This $\mathbb{Z}_2$ symmetry is known as the fermion parity operator $(-1)^F$.

In a BPS kink background of $N=(1,1)$ supersymmetry, there is a single
real fermion zero mode $\gamma$ which satisfying \[
\{\gamma, \gamma^\dag\}=1, \qquad \gamma=\gamma^\dag
\]
This zero mode can be thought as arising from the unbroken supercharge in this background.  Within the algebra $\mathrm{Cl}(1)$ generated by $\gamma$ alone,
the irreducible representation is one-dimensional \cite{Losev:2000mm} and $\gamma$ acts as a
$c$-number, $\gamma = \pm 1/\sqrt{2}$. However, if we further require that
the fermion parity $(-1)^F$ is a well-defined operator satisfying
$\{(-1)^F,\, \gamma\} = 0$ and $\bigl((-1)^F\bigr)^2 = 1$, the algebra is
enlarged to $\mathrm{Cl}(2)$, whose unique irreducible representation is
two-dimensional.\footnote{In \cite{Losev:2000mm}, it was argued that in the
physical limit $L\to\infty$, one of the two fermion zero modes delocalizes to
the spatial boundary, reducing the multiplet to a single state and rendering
$(-1)^F$ ill-defined. Whether the kink multiplet is one- or two-dimensional
thus depends on whether boundary-localized states are included in the physical
Hilbert space.} In this representation, $\gamma$ is no longer a $c$-number but
acts as a nontrivial operator on a two-dimensional Hilbert space spanned by
two states, which we denote by
\begin{equation}
|+\rangle, \qquad |-\rangle,
\end{equation}
and the zero mode acts as
\begin{equation}
\gamma |+\rangle \sim |-\rangle
\end{equation}

The kink multiplet therefore consists of two states:
\begin{equation}
\text{kink multiplet in } N=(1,1): \qquad 
\{\, |+\rangle,\; |-\rangle \,\}.
\end{equation}
Since $\{(-1)^F, \gamma\}=0$, it then acts on these two states as
\begin{equation}
(-1)^F |+\rangle = |+\rangle, \qquad
(-1)^F |-\rangle = - |-\rangle.
\end{equation}
if we assume $|+\rangle$ is bosonic.

\bigskip


\subsection*{2. The \texorpdfstring{$N=(2,2)$}{N=(2,2)} Case: One Complex Zero Mode}

The crucial difference from the $N=(1,1)$ case is that $N=(2,2)$
theories generally possess a continuous $U(1)$ symmetry that acts on
fermions.  And following previous discussion, we have 
\[
F=\sum_{E_n>0}a_n^\dag a_n-\sum_{E_n<0}b_m^\dag b_m-\frac{1}{2}\eta(0)+N_{0}
\]
The crucial point here is that the term of $\frac{1}{2}\eta(0)$ would contribute the fractional values to $F$ in the background of a BPS kink, where the other three terms contribute to the integer parts of $F$. To this end, let us consider the Diract opertor in this background 
\[
D=i\gamma^1\partial_x+m(x),\qquad m(x)=W''(\phi)\]
with this opeator, one can define the APS $\eta$ invaraint by looking at its eignevalue $\lambda$, which reads
\[
\eta(0)=\sum_{\lambda\neq 0} \text{sign}(\lambda)
\]
This is essentially the phase differnce across the kink. By denoting the compless mass as 
\[
m(x)=|m(x)|e^{i\theta(x)=W"(\phi_{kink}(x))}
\]
By acrossing the kink, the phase changes from $a$ vacuum to $b$ vacuum 
\[
\theta_a=\text{arg}(W^"(a)), \qquad \theta_b=\text{arg}(W^"(b))
\]
then the APS theory tells that 
\[
\eta(0)=\frac{1}{\pi}(\theta_b-\theta_a)+2k, \qquad k\in \mathbb{Z}
\]
So in general, the fermionic number operator $F$ in a kink background could have a non-trivial fractional part.

In summary,  one can define the fermion number as the generator of $U(1)$ symmetry, which exists for complex fermion in $N=(2,2)$ case, and $F$ can be fractional number. On the other hand, the Majarona fermion in $N=(1,1)$  does not have such a $U(1)$ symmetry, whereas it only has the fermion parity $Z_2$ symmetry $(-1)^F$, hence $F$ shall be integer.

\bibliographystyle{JHEP}

\newpage
\bibliography{refs.bib}

\end{document}